\documentstyle[multicol,aps,prl,eqsecnum,floats,epsf]{revtex}
\begin{document}
\draft
\title{
{\tenrm\hfill Preprint WIS-94/7/Jan-PH}\\
{\tenrm\hfill To appear in Phys. Rev. E}\\
STATISTICS OF A PASSIVE SCALAR ADVECTED BY A LARGE-SCALE
2D VELOCITY FIELD: ANALYTIC SOLUTION}
\author 
{M. Chertkov, G. Falkovich, I. Kolokolov \cite{kol}
and V. Lebedev \cite{leb}}
\address{Physics Department, Weizmann Institute of Science,
Rehovot 76100, Israel}
\maketitle

\begin{abstract}

\noindent

Steady statistics of a passive scalar advected by a random
two-dimensional flow of an incompressible
fluid is described in the range of scales between the correlation
length of the flow
and the diffusion scale. That corresponds to the so-called
Batchelor regime where the velocity is replaced by its large-scale gradient.
The probability distribution of the
scalar in the locally comoving reference frame is expressed via the
probability distribution of the line stretching rate.
The description of line stretching can be reduced
to a classical problem of
the product of many random matrices with a unit determinant.
We have found the change of variables that allows one to map
the matrix problem onto a scalar one and to thereby prove
the central limit theorem for the stretching rate statistics.
The proof is valid for any finite correlation time of the velocity field.
Whatever the statistics of the velocity field, the statistics
of the passive scalar (averaged over time locally in space)
is shown to approach gaussianity with increase in the Peclet number $Pe$
(the pumping-to-diffusion scale ratio).
The first $n<\ln Pe$ simultaneous correlation functions are expressed
via the flux of the square of the scalar and only one factor
depending on the velocity field: the mean stretching rate, which can be
calculated  analytically in limiting cases.
Non-Gaussian tails of the probability distributions at finite $Pe$ are found
to be exponential.\end{abstract}
\overfullrule=0pt

\section*{Introduction}
Description of a small-scale statistics of a passive scalar advected by
a large-scale solenoidal velocity field is a special problem in turbulence
theory. This problem was treated consistently from the very
beginning and some exact results have been obtained which is quite unusual
for a turbulence problem. Batchelor \cite{Bat} found exactly the form of
the double correlation function in the case of external velocity field
being so slow that it does not change during the time of the spectral transfer
of the scalar from the external scale to the diffusion scale. Then
Kraichnan \cite{Kra} obtained plenty of exact results in the opposite
limit of a velocity field delta-correlated in time. Those results are
valid for space of any dimension. We consider the two-dimensional problem
and show that it has some special features that allow one to develop an
analytical theory further and get some new qualitative and even
quantitative results.

We assume the velocity field to contain motions from some
interval of scales while the statistics of the scalar will be considered for
smaller scales beyond this interval. A steady
turbulence with a constant supply of the passive scalar is considered.
We wish to find the statistics of the passive scalar
$\theta$ in the convective interval of scales, i.e. for scales that are
less than both the
velocity correlation scale and the scale of the scalar supply $L$,  and larger
than the diffusion scale $r_{dif}$.
Since the scalar is a tracer in the velocity field,
then in order to find the statistics of the scalar, one should study the
statistics of line stretching first.

To get rid of the uniform sweeping and concentrate on the stretching
process we use a locally comoving reference frame \cite{Lvo,FalLeb,SS}.
All averaging is temporal. The source of the scalar is assumed to be
delta-correlated in the comoving frame.
As we show in Sect. I, the correlation functions of the scalar could be
expressed via the integrals of the correlation functions of the external
force along the trajectories of the fluid particles. For example, the value of
the pair correlation function $\langle\theta({\bf r}_1)\theta({\bf r}_2)
\rangle$ averaged over time in the locally comoving frame
is shown to be equal to the flux $P_2$ of $\theta^2$ multiplied by the
time $\tau_*$ that is necessary for the distance
between two fluid particles to increase from $R(0)=|{\bf r}_1-{\bf r}_2|$
to $R(\tau_*)=L$.
This makes it possible to reduce our problem to the equation
\begin{equation}
\dot{\bf R}(t)+\hat\sigma(t){\bf R}(t)=0 \,,
\label{E1}\end{equation}
where ${\bf R}$ is the two-dimensional
vector describing the separation of two points and
$\hat\sigma(t)$ is a  $2\times2$ matrix of the velocity
derivatives randomly varying with time and having some
{\em correlation time} $\tau$.
Due to incompressibility this matrix is traceless.
The main value of interest is the rate of line stretching
$\lambda(t)=t^{-1}\ln[R(t)/R(0)]$.

Studying the pair relative dispersion of the Lagrangian tracers is by itself
of great importance for describing spatially nonuniform situations such as
a pollutant spreading out into atmospheric turbulence. It is quite
well known \cite{Prod} how difficult it is to make some definite
statements about the statistics of $\lambda(t)$ and even about the mean
value $\bar\lambda=\lim_{t\rightarrow\infty}\lambda(t)$ which is usually
called the Lyapunov exponent. The difficulties are to do with the matrix
character of the equation (\ref{E1}). If this equation were scalar, then
it would be solved immediately: $\lambda(t)=-\int_0^t\sigma(t')dt'/t$. Due to
the central limit theorem, the value $\lambda(t)$ at $t\gg\tau$ would
have to have the Gaussian statistics with the mean $-\langle\sigma\rangle$
and with the variance $\int\langle\sigma(t')\sigma(0)\rangle\,dt'/t$
which decreases with time.

The situation is essentially the same as in the scalar case
if the matrix $\hat\sigma(t)$ is delta-correlated in time.
In this case, the rate of
stretching has a Gaussian statistics with the mean which
is determined solely by the pair correlation function:
$\bar\lambda=\int{\rm tr}\,\hat s(t')\hat s(0)dt'/4$ where
the so-called strain
$\hat s$ is the symmetric part of the matrix $\hat\sigma$.
A rigorous mathematical proof of the central limit theorem for $\lambda(t)$
in the delta-correlated case can be found in \cite{LaP}. The
statistics of the scalar is also easy to analyze in this case
(this was briefly exposed in \cite{FalLeb} and is
described in Sect. IIA and Appendix \ref{subsec:DC} in more detail).

Fortunately, the opposite limit of a long-correlated velocity field
(with the correlation time $\tau$ larger than the typical turnover time and
smaller than the transfer time $\tau_*$) can be analytically solved in 2D, too.
We have found $\bar\lambda={\rm Re}\Bigl\langle\sqrt{{\rm tr}\,
\hat\sigma^2/2}\,\Bigr\rangle$ -- see Sect. IIB. Note that in this case
both the symmetric and antisymmetric parts of $\hat\sigma$
determine $\bar\lambda$. Since the central limit theorem for $\lambda$
statistics
is readily proven in the slow case as well, it is tempting to assume that the
theorem is valid for an arbitrary value of $\bar\lambda\tau$.

The consideration of the  general (and most physically interesting)
case of a velocity field with the correlation time $\tau$ comparable with
the turnover time requires more sophisticated formalism.
The fact that the matrices $\hat\sigma(t_1)$
and $\hat\sigma(t_2)$ do not commute
prohibits a straightforward expression of
$R(t)$ as an exponent of some integral of $\hat\sigma$
(it could be written only via a time-ordered exponent).
The value $\lambda(t)$ is not an immediate object
of the central limit theorem. All correlation functions of $\hat\sigma(t)$
as well as its antisymmetric part
determining vorticity should be taken into account in the calculation of
$\bar\lambda$.

To analyze the general case, we suggest in Sect. IV a nonlinear substitution
(specific to 2D) that
enables one to write $R(t)$ as a plain exponent in terms of the new
variables.
A substitution of that kind was first introduced in the theory of
magnetism \cite{Kol} and it has proven useful in different problems
which reduce to time-ordered exponents of $2\times2$ matrices
\cite{Kol1}. In the general
case, those variables cannot be expressed analytically via the original
$\hat\sigma(t)$, yet some important properties could be established.
For example, we can prove the finiteness of the
correlation time of the random process in the new variables if $\tau$ is
finite.
This allows us to establish the central limit theorem for $\lambda(t)$, since
now it can be represented as an integral of some scalar quantity
(non trivially expressed through $\hat\sigma$) with a finite correlation time.
And what is probably more important, the substitution allows one to evaluate
the correlation time of the stretching rate fluctuations, which generally
different from the correlation time of the velocity field.
Our goal was also to find out whether some anomalies are
possible at $\bar\lambda\tau\simeq1$ that
prevent estimating $\bar\lambda$ by interpolation
between limiting cases.
For the particular case of the Gaussian velocity statistics with arbitrary
correlation time, this problem
is reduced in Sect.\ref{subsec:QM} to finding the ground state
of some not very complicated quantum mechanical system.
That made it possible to calculate $\bar\lambda$ numerically for the
different values of the correlation time and for the
different vorticity/strain ratios by using a pocket calculator.
The Lyapunov exponent has quite a simple
behavior which agrees with intuitive expectations:
$\bar\lambda$ monotonically grows with $\tau$ until
$\bar\lambda\tau\simeq1$ and then the dependence is saturated; $\bar\lambda$
monotonically decreases as a function of the vorticity/strain ratio. Such a
regular behavior
enables one to use, instead of numerics, a simple interpolation formula
explained in Sect.II. The formula expresses $\bar\lambda$ in terms of the
mean strain $S$, mean vorticity $\Omega$ and the correlation time $\tau$:
\begin{equation}
\bar\lambda= S\tanh{S\tau\over1+\Omega\tau}
\ .\label{E4}
\end{equation}
That formula satisfies all possible asymptotics and should give a reliable
estimate for any possible statistics of the velocity field.

Gaussianity of the stretching rate is an asymptotic property at
$t\rightarrow\infty$.
Since we are interesting in the stretching that provides for the spectral
transfer of
the passive scalar from $L$ to $r_{dif}$ then we always consider a finite time.
Measured at any finite time, the {\em probability distribution function}
(p.d.f.)
$P(\lambda)$ has generally non-Gaussian tails which we show to be exponential
for any velocity field (Sect. IIIA).

Extending the result of Furstenberg \cite{Fur} for
a finite $\tau$, we get the positiveness of $\bar\lambda$, so the
average
stretching is exponential in time for any velocity field. Consequently,
the stretching time $\tau_*$ logarithmically depends on distances
and so the pair correlation function:
\begin{equation}\langle\theta({\bf r}_1)\theta({\bf r}_2)
\rangle={P_2\over\bar\lambda}\ln{L\over|{\bf r}_1-{\bf r}_2|}\ .
\label{E3}\end{equation}
That expression is valid for $|{\bf r}_1-{\bf r}_2|=r_{12}\ll L$.
On the other hand, we neglect diffusion
which is possible for $r_{12}$ sufficiently large
to make the typical stretching time
$\bar\lambda^{-1}$ much less than the diffusion time $r_{12}^2/\kappa$.
Introducing $r_{dif}=(\kappa/\bar\lambda)^{1/2}$, the last condition could
be written as $r_{12}\gg r_{dif}$. The pair correlation function
$\langle\theta({\bf r}_1)\theta({\bf r}_2)\rangle$ logarithmically increases
as $r_{12}$ decreases until $r_{12}\simeq r_{dif}$; at smaller distances, the
growth saturates: with the logarithmic accuracy,
$\langle\theta({\bf r}_1)\theta({\bf r}_2)\rangle\approx\langle\theta^2\rangle=
P_2\bar\lambda^{-1}\ln(L/r_{dif})$.

Section IIIB describes the statistics of the passive scalar.
The p.d.f. $P\{\theta\}$ is generally
non-Gaussian for any finite Pe. It is interesting to find whether
the non-Gaussianity is related to the finiteness of the convective range or
it is inherent in the passive scalar dynamics and present even at the
limit of infinite Pe. Here we show that the non-Gaussianity is not an intrinsic
property of the advection but rather appears either due to a
non-ergodic nature of the
flow or as a result of an interplay between the advection and diffusion.
By a straightforward calculation, we obtain the expressions for the high-order
correlation functions (this was briefly exposed in \cite{FalLeb}).
The logarithmic case, we are dealing with, is substantially
simpler than cases with power-like correlation functions usually encountered in
turbulence problems.

One should distinguish between the statistics
of the products and of the differences of $\theta$ in different spatial points.
The statistics of the products is especially simple in our logarithmic regime.
We show that, as long as all the distances between the points are much less
than $L$,
the mean value of the product of $\theta$-s in $n$ points
is given by the reducible parts
(i.e. is expressed via the pair products) until $n<\ln(L/r)$ where $r$ is
either the
smallest distance between the points or $r_{dif}$ depending on what is larger.
In particular, $\langle\theta^{2n}\rangle=(2n-1)!!\langle\theta^{2}\rangle^n$
for
$n\ll\ln (L/r_{dif})=\ln Pe$. The reason for such Wick decoupling of the first
$n$ moments is simply the fact that reducible parts contain more large
logarithmic
factors than non-reducible parts do.
That means that $P\{\theta\}$ has a Gaussian core
(that describes the first moments) and
remote non-Gaussian tails (that describe moments with $n\gg\ln Pe$). We show
that
the tails are exponential (see also \cite{SS}). Let us emphasize that this is
true
for the statistics of the products taken at the points that are separated by
however small distances.

In contrast, the statistics of the differences depends on whether
the distance is in the convective interval ($r\gg r_{dif}$) or in the
diffusion interval ($r\ll r_{dif}$). The point is that the mean values
$\langle(\theta_1-\theta_2)^n\rangle$ are logarithmic only
for the distances $L\gg r_{12}\gg r_{dif}$. For example,
$\langle(\theta_1-\theta_2)^2\rangle=2\langle\theta^{2}\rangle-2\langle
\theta_1\theta_2\rangle=2P_2\bar\lambda^{-1}\ln(r_{12}/r_{dif})$. Let us
consider
$\langle(\theta_1-\theta_2)^4\rangle=
2\langle\theta^{4}\rangle+6\langle\theta_1^2\theta_2^2\rangle-4
\langle\theta_1\theta_2(\theta_1^2+
\theta_2^2\bigr)\rangle$. Substituting here the mean values of the products,
one finds
$3\bigl\langle(\theta_1-\theta_2)^2\bigr\rangle^2$.
Thus in the convective interval Wick decoupling is valid
while at the diffusion interval the reducible contributions cancel.
As a result, the statistics of the
differences is getting more Gaussian as $r_{12}/L$ decreases yet as the
distance
$r_{12}$ approaches the diffusion scale  the
statistics again starts to deviate from Gaussian. In the diffusion interval,
one can easily find $\langle(\theta_1-\theta_2)^2\rangle=P_2r_{12}^2/4\kappa$.
To determine higher moments and describe the statistics of the differences
in the diffusion interval one needs further studies.

As far as the dependence on the velocity field is concerned,
the Gaussian parts of the scalar distributions [giving lower moments
with $n<\ln(L/r)$] are determined solely by the value $\bar\lambda$. It is
remarkable that the non-Gaussian exponential tails are also universal and are,
up to
a dimensionless factor (depending on $\hat\sigma$ statistics), determined
by the mean value $\bar\lambda$ and variance $\Delta$.
Note also that (\ref{E3}) is true at small scales for any large-scale
turbulent velocity field even one containing some long-living
vortices that could trap the passive scalar for a long time, locally
suppressing stretching.
Note that we obtain the result on asymptotic Gaussianity by only temporal
averaging. If there are separate space regions with different
values of the pumping or the mean stretching rate (the flow is non-ergodic)
and if one desired to average with respect to
such a super-ensemble then gaussianity is lost while the logarithmic
dependencies of the correlation functions persist. Yet a single-probe
measurement
should reveal the probability distribution function with a Gaussian core and
exponential tails.

The subject of a local Gaussianity of the passive scalar was considered
as a bit confusing
due to the existence of an infinite number of integrals of motion
$I_n=\int\theta^n(r)\,d^2r$ in the undamped unforced case.
It was supposed \cite{Kra} that the fluxes $P_n$ of $I_n$ should determine
high-order correlation functions so that the statistics should depend on
the pumping that determines $P_n$. It is not the case for
logarithmic correlation functions.
We shall show in Sect. IIIC that the fluxes of $I_n$ for $n>2$
are not constant in the convective interval. This is happened due to an effect
of ``distributed pumping'' \cite{Fal}: high-order integrals are pumped even
in the convective interval of scales due to nonzero correlation functions
of the force with lower powers of $\theta$. As a result, the whole set of
the correlation functions (until $n\simeq \ln Pe$)
is solely determined by the pair correlation
function that is by the values $P_2$ and $\bar\lambda$.

The paper is organized as follows: the first part (Sects. I-III) contains
all physical statements supplied by a moderate mathematical formalism; those
who need more mathematical strictness and quantitative precision can find
those at the second part that includes Sect. IV and Appendices.

\section{Formulation of the problem}

We formulate the problem following Ref. \cite{FalLeb}.
The advection of the scalar field $\theta(t,{\bf r})$
is governed by the following equation
\begin{equation}
\dot{\theta}+u_{\alpha}\nabla_{\alpha} \theta =\phi +\kappa\Delta\theta,
\label{i1}
\end{equation}
where ${\bf u}(t,{\bf r})$
is the external velocity field and $\phi(t,{\bf r})$ is the external source
which we assume to be random functions of $t$ and ${\bf r}$.
We assume that the source $\phi$ is correlated on a scale $L$.
It means e.g. that the pair correlation function of the source
$\langle\phi({\bf r}_1,t_1)\phi({\bf r}_2,t_2)\rangle=\Xi(t_1-t_2,r_{12})$
as a function of the argument $r_{12}$
decays on the scale $L$. The same behavior is assumed for high-order
correlation functions of the source. The velocity field might be multi-scale,
its smallest scale is assumed to be larger than or of the order of $L$.
We consider a statistical steady state with a source provided by $\phi$ and a
small-scale sink due to diffusion with diffusivity $\kappa$ (note that most
of the below results  are independent of the particular form of the sink).
The mechanism of spectral transfer from
the pump to the sink due to stretching by an inhomogeneous
velocity field is clearly explained in \cite{Bat,Kra}.

To eliminate homogeneous sweeping, we pass to the  reference
frame locally comoving with the fluid at some point ${\bf r}=0$
(here and below ${\bf r}$ denotes radius-vector of a point in the
comoving frame). It corresponds to introducing the quasi-Lagrangian
velocities ${\bf v}(t,{\bf r})$ related to the initial Eulerian ones as
${\bf u}(t,{\bf r})={\bf v}\Bigl(t,{\bf r}-\int^t {\bf v}(0,t')dt'\Bigr)$.
We aim at finding simultaneous correlation functions of $\theta$ which
are the same for both sets of variables.
The equation (\ref{i1}) takes the form
\begin{eqnarray} 
\dot{\theta} +(v^{\alpha}-v_0^{\alpha})\nabla_{\alpha}\theta =
\phi+\kappa\Delta\theta,
\label{i4} 
\quad {\bf v}_0={\bf v}(t,0).
\end{eqnarray}
We will study the correlation functions of the scalar at different spatial
points separated by the distance that is smaller than the correlation
length $L$. We thus consider the points in space with the distances
from the zero point (where the sweeping is excluded)
to be also much smaller than the typical scale of velocity variations.
It allows one to expand the difference $v^{\alpha}({\bf r})-v^{\alpha}(0)=
\sigma^{\alpha\beta} r^{\beta}$. Here
$\hat{\sigma}(t)$ is the
matrix of the velocity derivatives which contains symmetric (strain)
and antisymmetric (vorticity) parts:
\begin{equation}
\hat{\sigma}=\hat{s}+\hat{\sigma}_{a}=
\left( \begin{array}{cc} a & b \\
b & -a \end{array} \right)+\left( \begin{array}{cc} 0 & c \\
-c & 0 \end{array} \right)\ .
\label{i41}
\end{equation}
We assume that $a(t),b(t)$ and $c(t)$ are independent
random processes with zero means and $\langle a^2\rangle=\langle b^2\rangle$.
The strain is determined by $a,b$ while the vorticity by $c$.

The resulting equation for $\theta(t,{\bf r})$ is
\begin{equation}
\dot{\theta}(t,{\bf r}) +\sigma_{\alpha\beta}(t) r_{\beta}\nabla_{\alpha}
\theta(t,{\bf r}) =\phi(t,{\bf r})+\kappa\Delta\theta(t,{\bf r})\ .
\label{i6}
\end{equation}
A formal solution of (\ref{i6}) is written in Appendix \ref{sec:dif}.
It is shown there, that as long as we are going to consider
the correlation functions
of the scalar at the the distances large comparing to
$r_{dif}\equiv\sqrt{\kappa/\bar\lambda}$, the diffusion term
could  be neglected.
We thus omit for a while the term $\kappa\Delta\theta$,
we shall bring that term back in Sect. IIIC in considering
the conservation laws and in Appendix \ref{sec:dif} in considering
one-point statistics of $\theta(t,{\bf r})$.
A formal solution of the equation
$$\dot{\theta}(t,{\bf r}) +\sigma_{\alpha\beta}(t) r_{\beta}\nabla_{\alpha}
\theta(t,{\bf r}) =\phi(t,{\bf r})$$
could be written as follows
\begin{equation}
\theta(t,{\bf r})=
\int_0^{\infty}dt'\phi\Bigl(t-t',\hat W(t,t-t'){\bf r}\Bigr)\,,
\label{i8} \end{equation}
where the matrix $\hat{W}$ should satisfy the following equation
\begin{equation}
\partial_t\hat{W}(t,t')+\hat{W}(t,t')\hat{\sigma}(t)=0\ .
\label{i7} \end{equation}
The initial condition is $\hat{W}(t,t)=\hat{1}$. The solution of
(\ref{i7}) can be written in the following form
\begin{equation}
\hat W(t,t') = \tilde T \exp \biggl( -\int _{t'}^t
dt_1 \, \hat\sigma(t_1) \biggr) \,,
\label{ttt}
\end{equation}
where $\tilde T$ designates the anti-chronologically ordered exponent.

By using (\ref{i8}), the  correlation functions of the scalar
can be rewritten in terms of the known correlation functions of the
pumping. The correlation time  of the pumping in the Lagrangian
frame $\tau_\phi^L$ will be assumed to be much less than $\bar\lambda^{-1}$
(which,
in a typical case, is of the order of the turnover time of the vortices of size
$L$). Since the pair correlation function of the source in
the frame moving with the fluid
$$\Xi\Biggl(t_1-t_2,\biggl|{\bf r}_1-{\bf r}_2+\int_{t_1}^{t_2}{\bf v}(0,t)
\,dt\biggr|\Biggr)$$
decays as a function of $(t_1-t_2)$ due to both arguments,
then $\tau_\phi^L=\min\{\tau_\phi^E,L/V\}$ where $\tau_\phi^E$ is the
correlation
time
in the Eulerian frame and $V$ is the mean turbulent velocity.
Further consideration will be valid if either the Eulerian correlation time of
the pumping is much less than the turnover time of $L$-eddies or the mean
turbulent velocity is much larger than the typical velocity of $L$-eddies.
The latter could be the case if, due to the inverse energy cascade in 2D
(at the scales larger than $L$),
the mean turbulent velocity $V$ that sweeps the scalar is determined by
the largest scale while the strain and the vorticity (that determine
$\bar\lambda$)
are determined by
the eddies with the scale $L$.
We thus write $\langle\phi({\bf r}_1,t_1)\phi({\bf r}_2,t_2)\rangle
=P_2\xi_2(r_{12})\delta(t_1-t_2)$ where the function $\xi_2(r_{12})$
describes spatial correlations of the pumping; $\xi_2(0)=1$.
The constant $P_2$ has physical meaning of the production rate of $\theta^2$.

The simultaneous pair correlation function of the scalar is written as follows
\begin{equation}
\langle\theta ({\bf r}_1)\theta ({\bf r}_2)\rangle=
P_2\biggl\langle\int_{0}^{\infty}dt \,
\xi_2\bigl(\mid\hat{W}(0,-t)({\bf r}_1-{\bf r}_2)\mid\bigr)
\biggr\rangle_\sigma\ .
\label{i10}
\end{equation}
where $\langle\ldots\rangle_\sigma$ denotes the average over
the statistics of $\sigma$. Averaging with respect to the
both external velocity and external source is implied
on the left-hand side of (\ref{i10}). We suggest that the statistics is
homogeneous in time, it enabled us to take the $\langle\theta\theta\rangle$
correlator in (\ref{i10}) at zero time.
The only unknown function in this expression is $\hat{W}(0,-t)$.
The function $\xi_2$ is determined by the statistics of the source $\phi$.
We can put simply $\xi_2(x)=1$ for $x<L$ and $\xi_2(x)=0$ for $x>L$ [the
account of any shape of $\xi_2(x)$ will give the same results with a
logarithmic
accuracy]. In this case,
\begin{equation}
\langle\theta ({\bf r}_1)\theta ({\bf r}_2)\rangle=
P_2\langle t(r_{12})\rangle_\sigma =
P_2\tau_*(r_{12})\ ,
\label{i11}
\end{equation}
where $t(r_{12})$ is the time necessary for two points to
increase their distance from $r_{12}$ to $L$ under the action of the
transfer matrix $\hat{W}(0,-t)$ and $\tau_*$ is $t(r_{12})$
averaged over the statistics of $\hat\sigma$. It is quite natural that
the pair correlation function is proportional to the time of separation:
imagine the ``heater'' $\phi$ of the size $L$, then
the values $\theta_1$ and $\theta_1$
of the ``temperature'' are correlated until the cold
fluid comes from outside into one of the points.

Since it is the modulus $|\hat W{\bf r}|$ that enters (\ref{i10})
then it is useful to represent $\hat\sigma$ as a sum of its symmetric part
$\hat s$ responsible for the stretching and antisymmetric one
$\hat\sigma_a$ that describes rotation. We get $\hat{W}$ in the form of
the product $\hat{W}=\hat{W}_s\hat{W}_a$, where the multipliers  satisfy
the following separate equations
\begin{equation}
\dot{\hat{W}}_a+\hat{W}_a\hat{\sigma}_a=0, \qquad
\dot{\hat{W}}_s+\hat{W}_s\hat{\tilde s}=0
\label{i72}
\end{equation}
with $\hat{\tilde s}=\hat{W}_a \hat s\hat{W}^T_a$.
The first equation is immediately integrated
\begin{equation}
\hat{W}_a(t,t')=\exp\biggl({-i\hat{\sigma}_y
\int_{t'}^{t}cd\tilde{t}}\biggr)
\ , \ \
\hat{\sigma}_y=\left( \begin{array}{cc} 0 & -i\\
i & 0 \end{array} \right).
\label{i73}
\end{equation}
Since $|\hat W_a{\bf r}|=r$ then it is $\hat W_s$ that should be substituted
in (\ref{i10}) but $\hat W_s$ actually depends on the effective strain
$\hat{\tilde s}$ which is determined by the whole set $a(t)$, $b(t)$
and $c(t)$.

\section{Limiting cases of a rapid and slow strain}

Here we show how the equation (\ref{i7})
can be directly solved in the limiting cases.
It is more convenient for us to consider the equivalent problem
of the behavior of the vector
\begin{equation}
{\bf R}(t)=\hat W(0,-t){\bf r}
\label{Rwr} \end{equation}
which determines e.g. the pair correlation function of $\theta$
via (\ref{i10}). Differentiating ${\bf R}(t)$ and using (\ref{i7})
we get (\ref{E1}). Here we aim at finding the probability distribution
function $P(t,\lambda)$ for $\lambda=(1/t)\ln(R(t)/r)$.

\subsection{Shortly correlated velocity field}

If the correlation time $\tau$ of the velocity derivatives is much less
than the turnover time (which is of the order  of the inverse mean strain
or vorticity) then
the random matrix $\hat\sigma$ can be considered as delta-correlated in time.
This case was first solved by Kraichnan \cite{Kra}
for a sparse distribution of sheets of the passive scalar.
One can discretize the equation (\ref{i7}) and represent the solution as
a product of random matrices from $SL(2,R)$. For a delta-correlated case,
those matrices are independent which allowed Furstenberg \cite{Fur}
to prove the positivity of the Lyapunov exponent and La Page \cite{LaP,Prod}
to prove  a central limit theorem: For all generic initial vectors ${\bf r}$,
the function
$$\Delta(t,{\bf r})={1\over t}\biggl\langle\biggl(\ln {R(t)\over r}-
t\bar\lambda\biggl)^2
\biggr\rangle$$
converges with increasing $t$ to a constant $\Delta>0$ independent of
$t$ and ${\bf r}$; the value $[\ln (R/r)-t\bar\lambda]/\sqrt{\Delta t}$
converges in distribution to a Gaussian standard random variable.

Our formalism allows for a compact description of that case. Let us
calculate the averages
$\langle R^{2n}(t)\rangle=
\langle ({\bf r}^T\hat W^T(0,-t)\hat W(0,-t){\bf r})^n \rangle$.
Substituting here the expression (\ref{ttt}) for $\hat W(0,-t)$
expanded in powers of $\hat\sigma(t')$ and
calculating the difference between the
instants $t$ and $t+\Delta t$ one gets
\begin{eqnarray}
& & \Delta\langle R^{2n}(t)\rangle=n\biggl\langle {\bf R}^T(t)
\int\int dt_1dt_2 2\hat s(t_1)\hat s(t_2)
{\bf R}(t)R^{2n-2}
\biggr\rangle\nonumber\\
& &+{n(n-1)\over2}\biggl\langle\biggl({\bf R}^T(t)
\int dt_1 2\hat s(t_1){\bf R}(t)\biggr)R^{2n-4}
\biggl({\bf R}^T(t)
\int dt_2 2\hat s(t_2){\bf R}(t)\biggr)\biggr\rangle\ ,
\nonumber \end{eqnarray}
where $t_1$ and $t_2$ run between $-t-\Delta t$ and $-t$.
We have chosen $\Delta t$ sufficiently small to allow the expansion in
$\tilde T$-exponents and sufficiently large to neglect the correlations
between $s(t_1)$ and $s(t_2)$ so that $\tau\ll\Delta t\ll S^{-1}$.
The terms determined by irreducible correlation functions
of $\hat \sigma$ are small in $\tau / \tau_*$, since they contain additional
restrictions for the region of integration over times $t_i$.
Accounting for the tensor structure
$\langle s_{\alpha\beta}(t_1) s_{\gamma\delta}(t_2)\rangle=D_s\delta(t_1
-t_2)(\delta_{\alpha\gamma}\delta_{\beta\delta}+\delta_{\alpha\delta}
\delta_{\beta\gamma}-\delta_{\alpha\beta}\delta_{\gamma\delta})$
one gets the equation
$\Delta\langle R^{2n}(t)\rangle=\langle R^{2n}(t)\rangle\Delta tD_s2n(n+1)$
where
\begin{eqnarray}
D_s&&={1\over8}\int\,{\rm tr}\,[\langle\hat\sigma(t)\hat\sigma(0)
\rangle+\langle\hat\sigma(t)\hat\sigma^T(0)\rangle]\,dt\nonumber\\ &&
={1\over4}\int{\rm tr}\,\langle\hat s(t)\hat s(0)\rangle\,dt
\ .\label{fast}\end{eqnarray}
Its solution
\begin{equation}\langle R^{2n}(t)\rangle=r^{2n}e^{D_s2n(n+1)t}\ .\label{Rn}
\end{equation}
exactly corresponds to the average
$$\langle R^{2n}(t)\rangle=r^{2n}\int P(t,\lambda)e^{2n\lambda t}\,d\lambda$$
with the Gaussian probability function
\begin{equation}
P(t,\lambda)=\sqrt{t/2\pi\bar\lambda}\exp[-(\lambda
-\bar\lambda)^2t/2\bar\lambda] \,,
\label{Gauss}\end{equation}
leading to $\bar\lambda=\Delta=D_s$.
We would like to stress
that in the white noise limit the
formula (\ref{Gauss}) is exact for arbitrary $t$.
The alternative method to get this
result can be found in \cite{ChFyKo}, the development of that method is
used in Sect. IV.

Above calculations are valid until $n\bar\lambda\Delta t\ll1$.
Since it should be $\Delta t>\tau$ then (\ref{Rn}) is valid for the
moments with $n<(\bar\lambda \tau)^{-1}$. Therefore the probability
distribution function $P(t,\lambda)$ has non-Gaussian corrections
due to a finiteness of the ratio $t/\tau$. In a formal limit of a
delta-correlated strain, $P(t,\lambda)$ is Gaussian everywhere for
any finite time $t$.

\subsection{Slow stretching and 1d localization}

It is worth noting that the vorticity gave no contribution
in the delta-correlated case. This was already
clear from (\ref{i72},\ref{i73}) since the correlation functions of
$\hat{\tilde s}$ coincide with those of $\hat s$ in this case.
For a finite correlation time, the vorticity plays an essential role
suppressing stretching due to the rotation of a fluid element with respect to
the axis of expansion and contraction.
Let us illustrate this by considering the simplest case of a time-independent
velocity field. Following Batchelor \cite{Bat} we consider a solution
$\theta({\bf r},t)=A\sin[{\bf k}(t){\bf r}]$ and get $k(t)=k(0)\exp\Bigl(t
\sqrt{a^2+b^2-c^2}\Bigr)$. That formula is valid if $\tau\gg t$.
Everywhere in this paper we are interested in the opposite case
$t\gg\tau$ when a universal statistics could appear.

An account of variations with time (even
slow ones) is quite difficult in a general case. Still,
the case of a slow velocity field also can be exactly solved.
We assume that the matrix $\hat\sigma(t)$
does not change substantially during a typical turnover time.
We differentiate the equation (\ref{E1})
$\dot{\bf R}=-\hat\sigma{\bf R}$ with respect to time and neglect
$\dot\sigma$ in comparison with $\sigma^2$. And here a little miracle happens:
because of incompressibility, the matrix $\hat\sigma$ is traceless so that
its square is proportional to the unit matrix in the 2D case. We thus come to
the scalar equation instead of the matrix one, this scalar equation can be
written in the form
\begin{equation}
\partial_t^2(R_x+iR_y)=(a^2+b^2-c^2)(R_x+iR_y)\ ,\label{loc1}
\end{equation}
One can consider (\ref{loc1}) as a Schr\"odinger
equation for a particle in a random potential $U=a^2+b^2-c^2$;
time plays the role of coordinate. We thus encounter the
problem of the type considered in the 1D localization theory.
We should find the behavior of the solution of (\ref{loc1}) under
initial conditions given at $t=0$. It is similar to the computation
of a 1D sample resistivity in the Abrikosov-Ryzhkin formulation
(see \cite{AbRy} and \cite{Kol1} for more details).
Based on their results we can assert that for any relation between
$a,b$ and $c$ the modulus of
$R_x+iR_y$ grows unlimited with $t$ as $\exp(\bar\lambda t)$.
This exponential growth is described by the same exponent as
the exponentially decaying tails of a localized quantum $\psi$-function.

Our limit of a slow strain corresponds to a quasi-classical regime so that
$\bar\lambda$ can be calculated by using semi-classical approximation.
Classically allowed and forbidden regions should be considered separately.
If $U<0$ (the region is classically allowed) then $R_x+iR_y$ is the sum
of two oscillating exponents so that the value $R_j$ of
the modulus of $R_x+iR_y$ at the beginning  of this interval  is generally
of the order of its value $R_{j+1}$ at the end.
If $U>0$ (the region is classically forbidden)
then $R_x+iR_y$ is the sum of the increasing and decreasing
exponents. To estimate the ratio $R_{j+1}/R_j$ we can neglect the decreasing
exponent and find $R_{j+1}/R_j\sim\exp\biggl(\int \sqrt{U(t')}\,dt'\biggr)$,
the integral here is taken between the points
$t_j$ and $t_{j+1}$ where $U=0$. The typical distance
between these points is the correlation time $\tau$ which is assumed to be
much larger than both the inverse mean strain $S^{-1}$
(determined by $a,b$) and the inverse mean
vorticity $\Omega^{-1}$ (determined by $c$). That means that the exponent
determining
$R_{j+1}/R_j$ is large (it is just the reason enabling us to neglect the
decaying exponent, since it is exponentially small). We conclude that with an
exponential accuracy the ratio $(R_x+iR_y)(t)/(R_x+iR_y)(0)$ is determined
by the regions where $U>0$ and can be estimated as the product of $R_{j+1}/R_j$
for these regions. That means that the rate stretching
$\lambda(t)=\ln(R(t)/R(0))$ can be written as an integral
\begin{equation}
\lambda(t)={1\over t}{\rm Re}\int_0^t\sqrt{U(t')}\,dt'\,,
\label{slowt}\end{equation}
so that we have again a central limit theorem for the statistics of
$\lambda(t)$
at $t\gg\tau$: $P(\lambda)=\sqrt{t/2\pi\Delta}\,\exp[-(\lambda
-\bar\lambda)^2t/2\Delta]$.

The expression for the Lyapunov exponent follows from (\ref{slowt}):
\begin{equation}\bar\lambda={\rm Re}\bigl\langle\sqrt{a^2+b^2-c^2}
\bigr\rangle\label{slow} \,,
\end{equation}
which can be calculated for any given statistics of $a,b,c$.
The time intervals with a negative $U$ give no contribution to $\bar\lambda$
[in the main order in $(\bar\lambda \tau)^{-1}$]
since regions with predominant vorticity do
not change the modulus of {\bf R} in a slow case.

{}From (\ref{slowt}) we find the variance:
$$\Delta=\int\Bigl\langle{\rm Re}\sqrt{U(t')}\,
{\rm Re}\sqrt{U(0)}\,\Bigr\rangle_c\,dt'\ ,$$
where we use standard notation $\langle AB\rangle_c
\equiv \langle AB\rangle- \langle A\rangle \langle B\rangle$.
We conclude that $\Delta\sim S^2\tau$ if $S\sim\Omega$. Note that
$\bar\lambda$ is determined by simultaneous averages so it does not
depend of the correlation time $\tau$ (at given values of $S$ and $\Omega$),
while the dispersion $\Delta$ does depend on it. A rigorous treatment
in Sect.IVB confirms (\ref{slow}) -- see also \cite{CFKL}.
{}From (\ref{slow}) it follows that $\bar\lambda$ can be estimated as
$S$ for the case $\Omega\lesssim S$.

The case $\Omega\gg S$ deserves a
separate consideration since $\bar\lambda$ will be suppressed in this
case. By calculating different-time correlation functions of $\tilde s$ one can
see
that the correlation time of $\tilde s$ is either $1/\Omega$ or $\tau$
depending on
which value is less. For $\Omega\tau\gg1$ we get
the asymptotic law of decreasing the Lyapunov exponent
for the limiting case of a very strong vorticity:
\begin{equation}
\bar\lambda\sim S^2/\Omega\ .\label{sup}\end{equation}
To turn an estimate into
a quantitative answer one should specify the statistics of the
velocity field.
For Gaussian statistics the answer could be found in \cite{CFKL} and
Section IV B for any $S,\Omega$.

To estimate the value of the Lyapunov exponent
$\bar\lambda$ in a simplified way,
one may construct an interpolation formula for $\bar\lambda$ in terms of
the mean strain $S$, mean vorticity $\Omega$ and correlation
time $\tau$. The value of $\bar\lambda$
for the fast case is equal to $D_s$ given by
(\ref{fast}) which can be estimated as $S^2\tau$. Taking into account
also the fact (which will be proved in Sect.IV) that the asymptotics
(\ref{slow}) in the slow limit
is approached exponentially in $S\tau$ we come to (\ref{E4}).
The concrete values of $S$, $\Omega$ and $\tau$ entering this expression
can be estimated from the pair correlation functions of $a,b,c$
up to numerical factors of the order unity (depending
on the statistics of $a,b,c$).

\section{Probability distributions: Gaussian bump and non-Gaussian tails}

We postpone the general proof of a central limit theorem for the stretching
rate
statistics until the next section. Here, assuming this theorem to be valid,
we study the probability distributions that appear for different quantities.
\subsection{Non-Gaussian tails of the probability distributions}
The central limit theorem assures us that
the probability distribution function for the
stretching rate $\lambda$ measured during the time $t$ that is much larger
than $\tau$ is as follows:
\begin{equation}
P(\lambda,t)=\sqrt{t/2\pi\Delta}\,\exp[-(\lambda
-\bar\lambda)^2t/2\Delta]\ . \label{Gauss1}\end{equation}
It is expressed via two parameters: Lyapunov exponent $\bar\lambda$ and
variance $\Delta$, which depend upon the statistics of the velocity
field. In the limiting cases, they can be expressed via
$S,\Omega$ and $\tau$ as it has been done in the previous section (see
also the next section); here we do not need their explicit form.
One can generally consider
$t$ and $\lambda$ as independent parameters in (\ref{Gauss1}).
$t\gg\tau$ being implied.
We will consider the limit of the large Peclet number $Pe$.
In this case, for separations $r_{12}$ taken in the
convective interval (from $r_{dif}$ to $L$) $\ln(L/r_{12})$ can be
treated as a large value.

We are interested in a particular case when $t$ is equal
to the time of passing the distance $R$ from $r_{12}$ up to
the external scale $L$. Then the values of
$\lambda$ and $t$ are simply related: $\lambda t=\ln(L/r_{12})$.
For such a relation, both distributions
\begin{equation}
P(\lambda)\propto\exp\biggl[-{(\lambda
-\bar\lambda)^2\ln{ (L/r_{12})}\over2\Delta\lambda}\biggr]\,,
\qquad P(t)\propto\exp\biggl[-{(\ln(L/r_{12}) /t
-\bar\lambda)^2t\over2\Delta}\biggr]\label{Gauss2}
\end{equation}
are generally non-Gaussian. For example, the time probability distribution
(that we need for evaluating the passive scalar statistics) is
close to Gaussian at $|t-\bar t|\ll\bar t=\ln{\rm (L/r_{12})}/\bar\lambda$.
At $t\gg\bar t$ that formula gives an exponential p.d.f.
$ P(t)\propto\exp(-t\bar\lambda^2/2\Delta)$.
Generally, $\lambda$ and the transfer time $t$ are related
in a more complicated way since $\lambda(t)$ is defined as some integral
over time of a fluctuating quantity. This influence pre-exponential
factors omitted in (\ref{Gauss2}) -- see the next subsection and
Appendix \ref{subsec:DC}.
There is another source of non-Gaussianity except the nonlinear relation
between $\lambda$ and $t$: the p.d.f. (\ref{Gauss1}) is itself
true only asymptotically as $t\rightarrow\infty$. At a finite $t$, p.d.f.
$P(\lambda,t)$ has non-Gaussian corrections that depend on the statistics
of $\hat\sigma$. In Appendix \ref{sec:Pc},
we show that the account of those corrections can add only a numerical
factor $c_2\simeq1$ in the exponent so that at $t\gg \bar t$
\begin{equation}
P(t)\propto\exp(-tc_2\bar\lambda^2/2\Delta)
\ .\label{Gauss3}\end{equation}
Note that this tail is not generally of the form $\exp(-t/\bar t)$.
For a delta-correlated case, $\Delta=\bar\lambda$ and $P(t)\propto
\exp(-\bar\lambda t/2)$. For a long-correlated case with $S\simeq\Omega$,
one has $\Delta\sim \bar\lambda^2\tau$ so that $P(t)\propto\exp(-ct/\tau)$
with the dimensionless coefficient $c$ that depends on the statistics
of $\hat\sigma$. For small $t\ll\bar t$ the probability sharply decreases:
$P(t)\propto\exp[-\ln^2(L/r_{12})/2t\Delta ]$.

\subsection{Statistics of the passive scalar}

The consideration of the
correlation functions of the passive scalar in the locally comoving reference
frame
(briefly presented
before in \cite{FalLeb}) will be based upon the representation ({\ref{i8})
giving the formal solution for $\theta$ in terms of the
pumping ``force'' $\phi$. To find a correlation function
of $\theta$ we should  first average  over
the statistics of $\phi$ and second over the statistics of $\hat\sigma$.
The result of the first averaging can be expressed in terms of the
correlation functions of $\phi$, an example is given by (\ref{i10}).
{}From that formula and (\ref{Gauss2}) one immediately gets the exponential
factor in the distribution function for the simultaneous product
$\langle\theta({\bf r}_1)\theta({\bf r}_2)\rangle_\phi=P_2t(r_{12})\equiv Q$
averaged over $\phi$ only:
\begin{equation}
{\cal P}(Q)\propto\exp\biggl[-{[\ln(L/r_{12})P_2/Q
-\bar\lambda]^2Q\over2P_2\Delta}\biggr],
\label{PdfQ}
\end{equation}
which is Gaussian near the maximum (within variance interval). The exponential
behavior far from the maximum given by (\ref{PdfQ}) is exactly correct only
for a delta-correlated $\hat\sigma$ [see (\ref{Psf1},\ref{Psf2})] otherwise
a numerical factor of order unity appears as well as in (\ref{Gauss3}).
To calculate the pre-exponential
factors in (\ref{Gauss2}) and (\ref{PdfQ}) it would be wrong to substitute
$\lambda t=\ln(L/r_{12})$ into (\ref{Gauss1}) -- such a substitution
would give the probability that erroneously counts trajectories that
reach $R=L$ at earlier time as well.
The contribution of the trajectories with nonmonotonous $R(t)$ into
pre-exponential
factor is substantial. To illustrate this, we calculate the whole ${\cal P}(Q)$
for the delta-correlated case -- see (\ref{Psf1},\ref{Psf2})
in the Appendix \ref{subsec:DC}.

In considering general correlation functions of $\theta$ we assume for
simplicity that the correlation functions
of odd products of $\phi$ are zero and therefore the statistics
of $\phi$ is characterized by the set of the irreducible correlation
functions $\Xi_2$, $\Xi_4$, \dots  of $\phi$. For example
\begin{eqnarray}
&&\langle \phi({\bf q}_1)\phi({\bf q}_2)
\phi({\bf q}_3)\phi({\bf q}_4)\rangle =
\Xi_4({\bf q}_1;{\bf q}_2;{\bf q}_3;{\bf q}_4)
\nonumber \\&&+
\Xi_2({\bf q}_1;{\bf q}_2)
\Xi_2({\bf q}_3;{\bf q}_4)+
\Xi_2({\bf q}_1;{\bf q}_3)
\Xi_2({\bf q}_2;{\bf q}_4)+
\Xi_2({\bf q}_1;{\bf q}_4)
\Xi_2({\bf q}_2;{\bf q}_3)
\label{U01} \,.
\end{eqnarray}
Here we designate ${\bf q}_i=(t_i,{\bf r}_i)$.
As we have explained previously, in the comoving reference frame,
one can treat the field $\phi$ as $\delta$-correlated
in time:
\begin{eqnarray}
\Xi_2({\bf q}_1,{\bf q}_2)\rightarrow&&
\delta(t_1-t_2)P_2\xi_2({\bf r}_1-{\bf r}_2)\,,
\nonumber \\
\Xi_4({\bf q}_1;{\bf q}_2;{\bf q}_3;{\bf q}_4)
\rightarrow&&\delta(t_1-t_2)\delta(t_1-t_3)\delta(t_1-t_4)
P_4\xi_4({\bf r}_1,{\bf r}_2,{\bf r}_3,{\bf r}_4)
\label{U02} \,,
\end{eqnarray}
etc. The quantity $P_4$ (introduced by analogy with $P_2$)
is the production rate of $\theta^4$ at the scale $L$.
Using those notations we can express the different-time pair correlation
function as follows
\begin{equation}
\langle\theta(t_1,{\bf r}_1)
\theta(t_2,{\bf r}_2)\rangle =
P_2\int\limits_{-\infty}^t dt'
\bigl\langle \xi_2 [\hat W(t_1,t'){\bf r}_1-
\hat W(t_2,t'){\bf r}_2] \bigr\rangle_\sigma
\label{U03} \,,
\end{equation}
where $t={\rm min}(t_1,t_2)$.
If $t_1=t_2$ then (\ref{U03}) reduces to (\ref{i10}).
For sufficiently small $r_1$ and $r_2$, the
integration time in (\ref{U03}) is large so that the absolute values of
$\hat W(t_1,t'){\bf r}_1$ and $\hat W(t_2,t'){\bf r}_2$
can be approximated by $\exp(\bar\lambda(t_1-t'))r_1$ and
$\exp(\bar\lambda(t_2-t'))r_2$ respectively. If $r_1\sim r_2$
then the argument of $\xi_2$ in (\ref{U03}) will be determined by
the largest of two times $t_1$ and $t_2$, say, $t_1$ (then $t=t_2$).
In this case the integration over $t'$ in (\ref{U03}) is
within the interval $t_1-\bar\lambda^{-1}\ln(L/r_1)<t'<t_2$. Therefore
with the logarithmic accuracy
\begin{equation}
\langle\theta(t_1,{\bf r}_1)
\theta(t_2,{\bf r}_2)\rangle =
P_2\bigl[\bar\lambda^{-1}\ln(L/r_1)-(t_1-t_2)\bigr]
\label{U04} \,,
\end{equation}
what implies $t_1>t_2$, $r_1\sim r_2$. The expression (\ref{U04})
is correct if $\bar\lambda^{-1}\ln(L/r_1)>(t_1-t_2)>\bar\lambda^{-1}$.
Note that the correlation function (\ref{U04}) does not depend only on the
difference $r_{12}$ since we lost
the homogeneity at passing to comoving reference frame. We see that
the correlation time of the scalar is logarithmically large in this frame
independently of the correlation time of $\sigma$. This is the manifestation
of the fact that $\theta$ is a Lagrangian invariant of the dynamics.

Consider now the higher-order correlation functions of $\theta$.
{}From (\ref{i8},\ref{U01},\ref{U02}) it follows
\begin{eqnarray}&&
\langle \theta_1\theta_2\theta_3\theta_4\rangle =
P_4\int\limits_{-\infty}^t dt'
\bigl\langle \xi_4 (\hat W(t_1,t'){\bf r}_1,
\hat W(t_2,t'){\bf r}_2, \hat W(t_3,t'){\bf r}_3,
\hat W(t_4,t'){\bf r}_4) \bigr\rangle_\sigma +
\nonumber \\&&
P_2^2 \biggl\langle
\int\limits_{-\infty}^{t_{12}}dt'
\xi_2 (\hat W(t_1,t'){\bf r}_1-
\hat W(t_2,t'){\bf r}_2)
\int\limits_{-\infty}^{t_{34}}dt''
\xi_2 (\hat W(t_3,t''){\bf r}_3-
\hat W(t_4,t''){\bf r}_4)
\biggr\rangle_\sigma +\dots
\label{U05} \,,
\end{eqnarray}
where the dots designate two additional terms originating from
$\Xi_2 \Xi_2$-products in (\ref{U01}), $t={\rm min}(t_1,t_2,t_3,t_4)$,
$t_{12}={\rm min}(t_1,t_2)$ etc. The same arguments as
previously show us that with the logarithmic accuracy the first
term in the right-hand side of (\ref{U05}) can be estimated as
$P_4\bar\lambda^{-1}\ln(L/r)$ and the second term in the right-hand
side of (\ref{U05}) is reduced to the product of the two pair
correlation functions (\ref{U04}). This product contains the
second power of the large logarithm and therefore the first
term in the right-hand side of (\ref{U05}) is negligible in
comparison with the second one. Thus we conclude that the main
contribution to the fourth-order correlation function
$\langle\theta_1\theta_2\theta_3\theta_4\rangle$
is determined by its reducible part which is the sum
$\langle\theta_1\theta_2\rangle\langle\theta_3\theta_4\rangle+
\langle\theta_1\theta_3\rangle\langle\theta_2\theta_4\rangle+
\langle\theta_1\theta_4\rangle\langle\theta_2\theta_3\rangle$.
The same assertion is obviously true for higher-order
correlation functions of $\theta$ until some order $n$
(see below). Therefore, the statistics of $\theta$ is Gaussian
with the
logarithmic accuracy which means that we calculated exactly only factors at
the terms with
the largest power of the logarithms while neglecting additive constants and
the terms with smaller powers of the logarithms.

Now we discuss the deviations from Gaussianity
due to a finiteness of Peclet number. Those deviations appear at
sufficiently large $\theta$
(or for sufficiently high orders of the correlators). We study them for
the simultaneous
correlation functions. Averaging  over the
statistics of $\hat\sigma$ can be replaced by
averaging over $\lambda$:
\begin{equation}
F_2=\langle\theta({\bf r}_1)\theta({\bf r}_2)\rangle=
P_2\int d\lambda P(t,\lambda)\lambda^{-1}\ln(L/r_{12})
\label{U06} \,.
\end{equation}
Here $P(t,\lambda)$ is p.d.f. of $\lambda$
on the time interval $t$ and $t\lambda=\ln(L/r_{12})$.
Since $P(t,\lambda)$ has the sharp maximum at $\lambda=\bar\lambda$
the integration over $\lambda$ in (\ref{U06}) gives (\ref{i10}).
Expressions analogous to (\ref{U06}) can be deduced for
higher-order simultaneous correlation functions of $\theta$. Consider
e.g. the fourth-order correlation function. If all separations
$r_{12}$, $r_{34}$,\dots are of the same order then with the
logarithmic accuracy
\begin{eqnarray}\FL
&&F_4=\langle \theta({\bf r}_1)\theta({\bf r}_2)
\theta({\bf r}_3)\theta({\bf r}_4)\rangle =
P_2^2\int d\lambda P(t,\lambda)\lambda^{-2}
\nonumber \\ &&
(\ln(L/r_{12})\ln(L/r_{34})
+\ln(L/r_{13})\ln(L/r_{24})
+\ln(L/r_{14})\ln(L/r_{23}))
\label{U07} \,.
\end{eqnarray}
The term $\int d\lambda P(t,\lambda)\lambda^{-2}$ in
(\ref{U07}) can be substituted by $\bar\lambda^{-2}$ which gives the sum of
the products of the pair correlation functions.
In the $2n$-th order correlation function the term
\begin{equation}\int d\lambda
P(t,\lambda)\lambda^{-n}\label{U077}\end{equation}
will arise.
It can be substituted by $\bar\lambda^{-n}$ if the number
$n$ is not very large. The largest value
of $n$ allowing for this substitution can be found by using (\ref{Gauss1}):
$n\lesssim(\bar\lambda/\Delta)\ln(L/r)$.
We conclude that the $2n$-th order correlation function of
$\theta$ is reduced to the product of the pair correlation
function up to the number $n\sim (\bar\lambda/\Delta)\ln(L/r)$
which is large due to the suggested large value of $\ln(L/r)$.
For higher $n$ the Wick theorem (that is the Gaussianity) is violated.

The crossover number $n\sim (\bar\lambda/\Delta)\ln(L/r)$
could be readily appreciated as the ratio of the transfer time
$\bar\lambda^{-1}
\ln(L/r)$ to the correlation time of the stretching rate fluctuation. The
latter is $\tau_\lambda\simeq \min\{\tau,\lambda^{-1}\}$ according to Sect.
\ref{subsec:FCT} while $\Delta\simeq\max\{\lambda,\lambda^{2}\tau\}$.
For the Gaussianity of the $n$-th correlation function, the time of mutual
correlations $n\tau_\lambda$ should be less than the transfer time.

To determine the value of $F_{2n}$ for
$n\gg (\bar\lambda/\Delta)\ln(L/r)$ it is worthwhile to pass
to the integration over $t$ using $t\lambda=\ln(L/r)$. Then
\begin{equation}
F_{2n}\simeq P_2^n (2n-1)!! \ln(L/r)
\int dt\,t^{n-2}P(t,\lambda)
\label{U08} \,,
\end{equation}
where all separations are assumed to be of the same order. For
large values of $n$ this integral is determined by large $t$. Substituting
$P(t)\propto\exp(-tc_2\bar\lambda^2/2\Delta)$
into (\ref{U08}) we find
$F_{2n}\propto (2n)! (P_2/2)^n \ln(L/r)\bigl(2\Delta/
c_2\bar\lambda^2\bigr)^{n-1}$.
This behavior can be described in terms of the probability distribution
function
\begin{equation}
P(\theta)\propto\ln(L/r)
\exp(-\sqrt{\bar\lambda^2c_2/P_2\Delta}\mid\theta\mid)
\label{U09} \,.
\end{equation}
We see that the exponent here does not depend on $\ln(L/r)$,
it enters (\ref{U09}) only as a factor. Therefore the
function (\ref{U09}) can be used to characterize
the large-$\theta$ tail of the single-point p.d.f.
$P(\theta)$, the only difference is that instead of $\ln(L/r)$ the
factor $\ln(L/r_{dif})=\ln Pe$ should be substituted into (\ref{U09}) ---
see Appendix \ref{sec:dif}.
This expression is valid at $\theta^2\gg\langle\theta^2\rangle
=(P_2/\bar\lambda)\ln Pe$. For most physically interesting case
$S\simeq\Omega\simeq\tau^{-1}$ one has $P(\theta)\propto\exp(-c_1
\sqrt{\bar\lambda/P_2}\mid\theta\mid)$ with some dimensionless coefficient
$c_1$ depending on the statistics of the velocity field. The basic statement
on the exponential tail of $P(\theta)$ agrees with that of Shraiman and
Siggia \cite{SS}, who examined the particular case of $\delta$-correlated
strain.

If the third-order correlation function of the pumping is nonzero then odd
correlation functions of the scalar are also nonzero. However,
they are logarithmically suppressed compared to even ones. That could be
established
by the procedure similar to what gave (\ref{U04}-\ref{U05}):
$\langle \theta_1^n\theta_2^{n+1} \rangle\propto \ln^n(L/r_{12})$.
The powers of the logarithm here are less than one would expect from the
scaling
$\theta\sim\ln^{1/2}$ prompted by the expressions for the even correlation
functions.

\subsection{Fluxes of the integrals of motion}
Initial equation (\ref{i1}) without pumping and diffusion conserves an infinite
sequence of the integrals $\int\theta^{2n}({\bf r})\,d{\bf r}$.
The way of pumping $\theta^2$ radically
differs from that of pumping high-order ($n>1$)
integrals \cite{FalLeb,Fal}.
For the steady flux of $\theta^2$-stuff in the convective interval
of scales $L\gg r_{12}\gg r_{dif}$,
one gets directly from (\ref{i1})
\begin{equation}
\langle\lbrack({\bf v}_1\bbox\nabla_1)+({\bf v}_2\bbox\nabla_2)\rbrack
\theta_1\theta_2\rangle=\langle \phi_1\theta_2+\phi_2\theta_1\rangle
+\kappa\langle \theta_1\Delta\theta_2+\theta_2\Delta\theta_1\rangle
\ .\label{flux2}\end{equation}
The first term in the right-hand side is constant at $r_{12}\ll L$ and it is
equal
to  $P_2$ which is the
pumping rate of $\theta^2$, while the second term is negligible for
$r_{12}\gg r_{dif}$. That means that the flux of $\theta^2$-stuff is constant
in the
convective interval.
For $\theta^4$ one gets similarly
$$\langle\lbrack({\bf v}_1\bbox\nabla_1)+({\bf v}_2\bbox\nabla_2)\rbrack
\theta_1^2\theta_2^2\rangle=
\langle \phi_1\theta_2^2\theta_1+\phi_2\theta_1^2
\theta_2+\kappa\theta_1\theta_2^2\Delta\theta_1+\kappa\theta_2\theta_1^2\Delta
\theta_2\rangle\ .$$
Besides the irreducible part
that is constant in the convective interval,
the correlator in the right-hand side necessarily contains the reducible parts.
The main contributions due to one-point means
$2\langle\theta^2\rangle\bigl(\langle\phi\theta\rangle+
\langle\theta\Delta\theta\rangle\bigr)$ are canceled because of the
conservation of
$\theta^2$ which requires $2\langle\phi\theta\rangle=-2\kappa
\langle\theta\Delta\theta\rangle=P_2$. In the different-point part we could
neglect
$\kappa\langle \theta_1\Delta\theta_2\rangle$ comparing to
$\langle \phi_1\theta_2\rangle$
as we did in considering (\ref{flux2}). We thus get
$\langle \phi_1\theta_2\rangle\langle\theta_1\theta_2\rangle=P_2\langle
\theta_1\theta_2\rangle$, which changes with $r_{12}$ as the pair
correlator. Since our pair correlation function is logarithmic, it grows
as $r_{12}$ decreases. This means that for sufficiently small $r_{12}/L$ one
can neglect the constant irreducible contribution determined by $P_4$ in
comparison with $P_2^2\bar\lambda^{-1}\ln(L/r_{12})$.
That is why all the fluxes for $1<n<\ln{\rm Pe}$
$$\langle\lbrack({\bf v}_1\bbox\nabla_1)+({\bf v}_2\bbox\nabla_2)\rbrack
\theta_1^n\theta_2^n\rangle\propto P_2^n\bar\lambda^{1-n}\ln^{n-1}(L/r_{12})$$
are expressed in terms of $P_2$ and are non constant in the convective
interval.
It is worth emphasizing that the reason for this is the presence of the
external action at any scale i.e. the absence of the convective
intervals for higher integrals. The flux change has nothing to do with
non conservation.

This simple consideration shows how an asymptotic Gaussianity appears for
logarithmic correlation functions and how the set of the correlation
functions appears to be independent of the influxes of higher integrals
of motion. In conclusion of this section we would like to repeat that
if the flow is non-ergodic and there are separate space regions with different
values
of the pumping then
at averaging of the correlation functions over space the Gaussianity
is lost (see
also \cite{KK}) while the logarithmic dependencies of the correlation
functions remain the same.

\section{Anzatz for T-exponent}

In Sect. II we have established the asymptotic Gaussianity of
the statistics of the stretching rate and found the Lyapunov exponent
at the limiting cases of rapid and slow strain.
Although velocity fields that produce such strain can exist,
the most interesting and widespread cases certainly correspond
to the strain correlation time that is of order of the turnover time.
Starting from this section we shall manage to prove a central limit theorem
and to find a way to calculate $\bar\lambda$ for an arbitrary $\tau$. For
pursuing the first aim we should find such a representation for
$\ln|\hat W(T,t){\bf r}|$ at large $T$ that can be presented as
an integral of a scalar function with a finite correlation time.

It is possible to extract from (\ref{i7},\ref{i72}) $\hat{W}_s(t)$ only as the
anti chronological time-ordering exponent (\ref{ttt})
not as some regular function of $\hat{\sigma}$. To calculate averages
over $\hat{\sigma}$ we can use the formalism of a path integral
\begin{equation}\langle f(\hat W)\rangle=\int{\cal D}
\hat\sigma\,\exp(-S\{\hat\sigma\})f(\hat W)
\label{path}\end{equation}
with an appropriate action $S\{\hat\sigma\}$
that determines the statistics of $\hat\sigma$.
Such a formalism enables one to pass from the variables $\hat\sigma$ to
another variables that give $\hat W$ as a regular function.
A similar problem -- transformation of time-ordered exponent of some
linear combination of spin $SU(2)$ operators --
has been solved by Kolokolov \cite{Kol} for an exact functional
representation of the partition function of quantum Heisenberg ferromagnet.
The main idea of the
anzatz is as follows: by using the commutation relations of the spin algebra,
to find out such new integration variables in the functional
integral, that $\tilde{ T}\exp$ becomes some regular function. Here,
we suggest a new modification of this anzatz. Besides the possibility
to establish the statistics of $\lambda$ for an arbitrary $\tau$, the
new representation will allow us to generate consistent perturbation theory
in both cases of a rapid and slow strain.

\subsection{Central limit theorem for an arbitrary correlation time}
First, we expand the $\hat{s}$ matrix (which is the symmetric part
of $\hat\sigma$) over the spin $2\times 2$ matrixes
$\hat{s}=a\hat{\sigma}_z+b\hat{\sigma}_x$.
Then we introduce a new basis of the spin algebra $\hat{\sigma}_y$,
$\hat{\sigma}_{\pm}$ with
\begin{equation}
\hat{\sigma}_{\pm}= \hat{\sigma}_z \pm i\hat{\sigma}_x=
\left( \begin{array}{cc} 1 & \pm i \\
\pm i &  -1 \end{array} \right)
\label{f21}
\end{equation}
that corresponds to the rotation of the quantization axis from the
usual position (parallel to the z-axis) to the new one -- parallel
to the y-axis. We choose, instead of $a(t),b(t)$ fields,
the new ones $\varphi^{\pm}\equiv(a\pm ib)/{2}$, representing $\hat{s}$
in a more compact form $\hat{s}=\varphi^{-}\hat{\sigma}_{+}+
\varphi^{+}\hat{\sigma}_{-}$.

Let us consider the matrix function given in the explicit form
\begin{equation}
\hat{A}(T,t)=
\exp\left[-\hat{\sigma}_-\Phi^+(t)\right]
\exp\left[-\hat{\sigma}_{+}\Phi^-(T)\right]
\exp\left[\hat{\sigma}_{y}\Phi^y(T)\right]
\exp\left[\hat{\sigma}_-\Phi^+(T)\right],
\label{f6}
\end{equation}
where $\Phi^{\pm},\Phi^y$ are the functionals
of the new dynamical fields $\psi^{\pm},\rho$:
\begin{equation}
\Phi^+(T)=\psi^+(T) \ , \
\Phi^-(T)=\int_t^T\psi^- e^{2\int_t^{\tilde{t}}\rho dt'}d\tilde{t} \ , \
\Phi^y(T)=\int_t^T\rho d\tilde{t}\,.
\label{f6a}
\end{equation}
We remind that the field $c$ characterizes the antisymmetric part of
$\hat\sigma$. Using the commutation relations of spin operators (\ref{f21})
one can show that the matrix function $\hat{A}$ obeys the differential equation
\widetext
\begin{equation}
\partial_T{\hat{A}}=\hat{A}
\bigl\{-\hat{\sigma}_+\psi^-+
\hat{\sigma}_-\bigl[4\psi^-(\psi^+)^2-
2\rho\psi^++{\dot\psi}^+\bigr]+\hat{\sigma}_y(-4\psi^-\psi^++\rho)\bigr\},
\label{f8}
\end{equation}
and the first factor in (\ref{f6}) ensures the boundary condition
$\hat{A}(t,t)=1$.
Comparing (\ref{f8}) with (\ref{i72})
we find that the substitution $\varphi^-=\psi^-$ and
\begin{eqnarray}
& & \varphi^+=-\dot{\psi}^+-2ic\psi^++4\psi^-(\psi^+)^2,
\label{f9}\\
& & \rho=4\psi^-\psi^+-ic,
\label{f10}
\end{eqnarray}
guarantees the coincidence of $\hat{W}_s$ and $\hat{A}$.
It allows us to obtain an explicit functional integral for any averages written
in terms of $\hat{W}_s$ by means of changing the variables ($\varphi^{\pm}
\rightarrow\psi^{\pm})$.
The transformation (\ref{f9}) contains the derivative of the field
$\psi^+$ with respect to time on its right-hand side. Therefore, it should
be supplied with some initial conditions.
Another point is that in the course of calculation
it is necessary to average some functions of the operator $\hat A(T,t)$ at
fixed time moment $T$ over the velocity statistics.
For a given $T$ it is convenient to fix the final
value of the field $\psi^+$ (see also \cite{Kol1,ChFyKo}): $\psi^+(T)=-1/2$.
The Jacobian of the map (\ref{f10})
is determinant of a triangle matrix and
depends on the choice of the regularization.
We choose here the same variant of discretization
of the map (\ref{f9}) as it was used before
in the papers \cite{Kol1},\cite{ChFyKo}
($\varphi^{\pm}_n=\varphi^{\pm}(t_n); n=1,...,M ; h=\frac{T-t}{M}
\rightarrow 0; t_n=t+h n ; M\rightarrow \infty $),
\begin{equation}
\varphi^-_{n}=\psi^-_n \ , \
\varphi^+_n=-\frac{1}{h}(\psi^+_{n+1}-\psi^+_n)-
ic_n(\psi^+_{n+1}+\psi^+_n)+
\psi^-_n(\psi^+_n+\psi^+_{n+1})^2,
\label{f13}
\end{equation}
that gives the following expression for the Jacobian
$${\cal D}\varphi^{\pm}={\cal J}[\psi^{\pm}]{\cal D}\psi^{\pm},
\;
{\cal J}=const \exp\left(4\int_t^{T} \psi^{+}\psi^{-} dt'-i\int_t^Tcdt'\right)
\,.$$
The matrix $\hat{A}(T,t)$ being multiplied on the initial vector
${\bf R}(T)=
\left( \begin{array}{c} 1\\0 \end{array}\right)$ produces the following
simple expressions for the squared vector (\ref{Rwr}):
\begin{equation}
{R}^2(T-t)=[\hat{A}(T,t){\bf R}(T)]^2=-2\psi^+(t)
\exp\biggl[{8\int_{t}^{T}(\psi^+\psi^--ic/4)dt'}\biggr].
\label{R2}
\end{equation}
Here we exploited the isotropy condition and chosen
${\bf R}(T)$
without any loss of generality.
The formula (\ref{R2}) immediately gives
a desired asymptotic (at large $T$)
expression for the stretching rate where only the real exponents contribute:
\begin{equation}
\lambda(T)={4\over T}\int_0^T\psi^+(t)\psi^-(t)dt
\ .\label{f17}
\end{equation}
We thus succeeded in representing the stretching rate as an integral of a
scalar function. Expression (\ref{f17}) allows us to prove positivity of
$\bar\lambda=\lim_{T\rightarrow \infty}\langle \psi^+(T)\psi^-(T)\rangle$
and the
central limit theorem for the statistics of $\lambda(t)$. At first glance,
the substitution (\ref{f9}-\ref{f10}) makes $\psi^+$ not to be complex
conjugated to $\psi^-$ so that $\psi^+\psi^-$
may be negative and even complex. It can be shown that
in calculating averages one can deform the integration contour in the plane
$\{\psi^+,\psi^-\}$ into the contour
$\bigl(\psi^+\bigr)^*=
\psi^-$ without encountering singularity \cite{CFKL}. We thus can conclude that
$\bar\lambda>0$. This could be also obtained from generalization
of the classical results
of Furstenberg \cite{Fur} to the case of finite
correlation time. To establish a central limit theorem one should prove that
the random process $\psi^+(t)\psi^-(t)$ has a finite
correlation time for a finite-correlated $\hat\sigma$. This will be shown
at Appendix B2. As time $t$ is getting much larger than the correlation
time of $\psi^+(t)\psi^-(t)$ then the statistics of
$\lambda(t)$ approaches a Gaussian one. Let us remind that $t$ is bounded
from above by $\tau_*=\bar\lambda^{-1}\ln {\rm Pe}$
so that the Gaussianity is an
asymptotic property of high Pe regime.

The representation (\ref{f17}) for the
stretching rate enables us to develop a consistent perturbation theory in both
limits of slow and fast strain. To this end, one should look for the value
$\bar\lambda=\lim_{T\rightarrow \infty}4\psi^+\psi^-$
substituting as $\psi^+$ the solution of the equation
(\ref{f9}) and averaging the result.
Extracting from the equations (\ref{f9}) $\psi^-$, one obtains
\begin{equation}
\varphi^++2ic\psi^+=-\dot{\psi}^++4\varphi^-(\psi^+)^2\ .
\label{cc}
\end{equation}
The right-hand side of the equation (\ref{cc}) consists of the two terms.
The first term is leading in the fast case while the second one is leading in
slow case. Then we can look for the
corrections to those leading terms what will produce two different
asymptotic series. In this way we obtain a set of recursion relations
enabling us to formulate asymptotic expansions for $\bar\lambda$
for the cases of the slow and fast fields. Such an iteration procedure can be
found in \cite{CFKL}, it confirms the simple approach of Sect. II.

\subsection{Gaussian strain with an arbitrary correlation time}
\label{subsec:QM}
Let us emphasize that up to now in this section we have not specified the
statistics of the velocity field. All the above statements have
universal character and do not depend on a detailed structure of
velocity statistics. Still, to get precise quantitative description
in the case of an arbitrary correlation time one should specify
the statistics. Let us choose for further investigation the case of
Gaussian statistics as the simplest (yet nontrivial) example.
Namely, we will consider the particular case of the following
Gaussian statistics of $\hat{\sigma}$:
\begin{eqnarray}
& & {\cal D} \hat{\sigma}(t)\exp(-S_0)=
{\cal D}a {\cal D}b {\cal D}c \exp(-S_0),\nonumber\\
& & S_0=\frac{1}{2D_s}
\int [a^2+b^2+\tau^2(\dot{a}^2+
\dot{b}^2)]dt+
\frac{1}{2D_a}
\int [c^2+\tau^2\dot{c}^2]dt
\label{i5}
\end{eqnarray}
The expression of $\sigma$ via the fields $a,b,c$ is
given by (\ref{i41}). In (\ref{i5})
we introduced two different values $D_s$ and $D_a$
characterizing respectively
the amplitudes of the strain (described by the fields $a$ and $b$)
and of the vorticity (described by the field $c$). The mean values
$S$ and $\Omega$ of the strain and the vorticity can now be defined as
\begin{equation}
S^2=D_s/\tau, \quad \Omega^2=D_a/\tau
\,. \label{som}
\end{equation}

To determine $\bar\lambda$ (or another averaged quantity) one should
calculate the functional integral with the measure that is obtained by
substituting (\ref{f9}) into (\ref{i5}).
Before doing it let us rewrite the measure by introducing auxiliary
fields $\xi^+$ and $\xi^-$:
\begin{eqnarray}
& &
{\cal D} \hat{\sigma}(t)\exp(-S_0\{a,b,c\})\Rightarrow
{\cal D}\varphi^{\pm} {\cal D}\xi^{\pm}{\cal D}c
\exp(-S_1\{\varphi,\xi,c\}),
\nonumber\\
& & S_1=\frac{2}{D_s}
\int\left[\varphi^{+}\varphi^{-}+\xi^+\xi^-+\varphi^+
\xi^-+\tau^2\ddot{\varphi}^-\xi^+\right]dt+\frac{1}{2D_a}
\int\left[c^2+\tau^2_a(\dot{c})^2\right]dt.
\label{2c1} \end{eqnarray}
Let us perform the substitution (\ref{f9})
and the following linear change of variables:
\widetext
\begin{equation}
\psi^-=D_s\pi^- \ , \ \ \psi^+=\eta^+ \ , \
 \xi^-=-D_s\pi^-+\frac{D_s}{2}\eta^- \ ,
\ \xi^+=-D_s\pi^+ \ , \ c=  \sqrt{D_s
 D_a}z
\nonumber
\end{equation}
The Lyapunov exponent is thus the average of
$4D_s \lim_{T\rightarrow \infty}\pi^-(T)\eta^+(T)$ with respect
to the new measure
\widetext
\begin{eqnarray}
& & N{\cal D}\pi^{\pm} {\cal D}\eta^{\pm}{\cal D}z
\exp(-S_3\{\pi,\eta,z\}),\nonumber\\
& &
 S_3=\int \left[2(D_s\tau)^2  \dot{\pi}^+\dot{\pi}^-
-\dot{\eta}^+\eta^-+2\pi^+\pi^--4\pi^-\eta^+-\pi^+\eta^-+4\pi^-(\eta^+)^2\eta^-
\right]dt'\nonumber\\
&&-
i\sqrt{\frac{D_a}{D_s}}\int (2\eta^+\eta^- -{1})z dt'
+\frac{1}{2}\int \left[z^2 +(\tau D_s)^2(\dot{z})^2 \right]dt' ,
\label{2c4}
\end{eqnarray}
where $t'=tD_s$.
The action (\ref{2c4}) is of the Feynman-Kac type \cite{FeHi}.
Our path integral is the matrix element of the quantum-mechanical
evolution operator $\exp(-\hat{H}_1T)$ where the Hamiltonian $\hat{H}_1$
corresponds to a system with three degrees of freedom:
\widetext
\begin{equation}
\hat{H}_1=-\frac{\Delta_r}{2(\tau D_s)^2}
+\frac{r^2}{2}-\frac{r}{2}[4e^{-i\vartheta}\hat{d}^-(1-
\hat{d}^+\hat{d}^-)+{e^{i\vartheta}\hat{d}^+}]-
\frac{1}{2\tau^2 D_s^2} \frac{\partial^2}{\partial z^2}+
\frac{z^2}{2}-2i\sqrt{\frac{D_a}{D_s}}\hat{d}^+\hat{d}^-z\ .
\label{2c5}
\end{equation}
Here $(r\cos\vartheta,r\sin\vartheta)=(\pi^++\pi^-,i\pi^--i\pi^+)$ ,
so that $\Delta_r$ is the Laplacian operator
in two-dimensional ${r}$-space, and $\hat{d}^+,\hat{d}^-$ are some
creation and annihilation operators with the usual commutation relation
$\{\hat{d}^-,\hat{d}^+\}=1$, corresponding to $\eta^-$ and $-\eta^+$
fields in the path integral over the measure (\ref{2c4}) accordingly.
Let us note, that essential difference between ``quantization'' procedures
for $\pi$ and $\eta$ fields stems from different structures of ``kinetic''
terms in the action (\ref{2c4}) (of Weyl and Wick types \cite{Ber,Per}
respectively). In general terms Wick's quantization procedure requires
to fix an ordering of creation and annihilation  operators in the
Hamiltonian. In our case, it is possible to show straightforwardly
that the regularization of the map (\ref{f13}) and corresponding
regularization of all the expressions which we used fixes the ordering
of operators in the Hamiltonian (\ref{2c5}) unambiguously.
However, there exists more simple way to check this statement.
Indeed, the energy of the ground state of the Hamiltonian (\ref{2c5}) must
coincide with the corresponding one of the harmonic oscillator.
But the only ordering satisfying this requirement is the one chosen in
(\ref{2c5}).

All the local in time averages with respect to the
measure (\ref{2c4}) are the respective averages over the ground state
of the Hamiltonian (\ref{2c5}).
So, to calculate the Lyapunov exponent one has to average
$4D_s\pi^-\hat{d}^-$ over the ground state of the Hamiltonian (\ref{2c5}).
In \cite{CFKL} this well-defined quantum mechanics is described in
details. The wave function of the ground state is found as an
expansion into the series over some polynomials,
where the expansion coefficients connect with each other linearly.
The system of linear equations for
the coefficients can be solved with any precision required.
The contribution of high-order polynomials
is negligible; the number of terms giving substantial contribution
grow with $\tau$ and is finite for a finite $\tau$.
We computed the Lyapunov exponent $\bar\lambda$
up to large enough correlation times ($\tau\approx 10 D_s^{-1}$).
Figures 1 and 2 summarize our numerical evidence for the Lyapunov
exponent as a function of the correlation time $\tau$ and the ratio
$D_a/D_s$. They show a good qualitative agreement with the
interpolation formula (\ref{E4}). It is worth noting that the Lyapunov
exponent $\bar\lambda$ is a monotonically growing function of the
correlation time $\tau$ at fixed mean values $S$ and $\Omega$ of the strain and
vorticity introduced by (\ref{som}) and $\bar\lambda$ is a monotonically
decreasing function of $\Omega$.

\section*{Conclusion}
We developed a theory for quite arbitrary temporal characteristics
of the turbulent velocity field. As far as the spatial requirement of
velocity being large-scale is concerned, it is not very restrictive
as well. If the energy spectrum $E(k)$ of the velocity field
decays at $kL>1$ fast enough to produce the main strain by the
scales of the order $L$, then the above theory is valid.
This is so, in particular, in a viscous-convective range at large Prandtl
number (viscosity to diffusivity ratio) \cite{LH} where the energy
spectrum $E(k)$ decays exponentially. What if we consider
the convective interval for the velocity as well so that the
spectrum is power-like $E(k)\propto (kL)^{-x}$?
For the distributions with $x>3$
\cite{Saff,Moff,Poly} the strain is large-scale and our theory
is applicable. Besides for steady turbulence
one can prove that $x\geq3$ (see e.g. \cite{FH}). As far as
turbulent vorticity cascade is concerned, it corresponds to
$x=3$. This case does not satisfy the applicability conditions of the above
theory and the calculation of the scalar p.d.f.
is still ahead of us. The $r$-dependencies of the scalar correlation functions
can be found nevertheless. Would $E(k)$ be exactly $k^{-3}$,
then all correlators were
proportional to
the integer powers of the logarithm as above; for instance,
$\langle\theta({\bf r}_1)
\theta({\bf r}_2)\rangle\propto\ln(L/r_{12})$ \cite{LH}. We know, however,
that the velocity spectrum is logarithmically renormalized
$E(k)\propto (kL)^{-3}\ln^{-1/3}(kL)$ \cite{Kraa,FalLeb}
so that both the effective strain and vorticity depend on the
scale and grow with $k$ as to
provide $\bar\lambda(k)\propto\ln^{1/3}(kL)$. In this case,
the correlation functions of the passive
scalar coincide with those of the vorticity:
$\langle
\theta^n({\bf r}_1)\theta^n({\bf r}_2)\rangle\propto\ln^{2n/3}(L/r_{12})$
\cite{FalLeb}.

\acknowledgements We strongly benefited from the discussions with
Y. Fyodorov and G. Samorodnitsky.
Useful remarks of R. Kraichnan, A. Polyakov, E. Siggia and
J. McWilliams are gratefully
acknowledged. This work was partly supported by the Rashi Foundation (G.F.),
by the Landau-Weizmann Program (V.L.) and by the Minerva Center for Nonlinear
Physics (I.K.)

\appendix
\section{Diffusion and correlation functions at small distances}
\label{sec:dif}

As long as one considers the correlation functions at sufficiently small
distances,
in particular one-point statistics, the account of diffusion or other
dissipative
mechanism is unavoidable. That makes no substantial difficulties yet some
formulas
are getting bulky. Here we show how the formalism similar to
(\ref{i8},\ref{i10})
could be applied to the complete equation (\ref{i6})
and prove that the one-point statistics
of $\theta$ is the same as the statistics of the different-point products in
the
convective interval: first $n<\ln Pe$ moments are
gaussian i.e. are determined by the value $\langle\theta^2\rangle$.

Let us look for a solution of (\ref{i6}) in the following form
\begin{equation}
\theta(t,{\bf r})=f(t,{\bf R})\equiv
f(t,\hat{W}(t,t_0){\bf r}),
\label{Dif2}
\end{equation}
where the evolution of the matrix $\hat{W}(t,t')$ is given by
(\ref{i7},\ref{ttt})
and $t_0$ is a fixed moment of time.
Then, the function $f(t,{\bf R})$ should satisfy the following equation
\begin{equation}
\dot{f}(t,{\bf R})=\phi (t,\hat{W}^{-1}{\bf R})+\kappa
W_{\alpha\gamma}W_{\beta\gamma}
\frac{\partial}{\partial  R_\alpha}\frac{\partial}{\partial  R_\beta}
f(t,{\bf R}).
\label{Dif3}
\end{equation}
This linear equation may be rewritten in the Fourier representation with
respect
to ${\bf R}$,
\begin{eqnarray}
&&\dot{f}_{\bf k}(t)=\tilde{\phi}_{\bf k}(t)-\kappa
(\hat{W}\hat{W}^{T})_{\alpha\beta} {\bf k}_\alpha{\bf k}_\beta f_{\bf k}(t),
\label{Dif4}\\&&
f_{\bf k}(t)\equiv \int d{\bf R} e^{i({\bf k R})}f(t,{\bf R}) ,\ \ \
\tilde{\phi}_{\bf k}(t)\equiv
\int d{\bf R} e^{i({\bf k R})}\phi (t,\hat{W}^{-1}{\bf R}).
\label{Dif5}
\end{eqnarray}
The dynamical equation (\ref{Dif4}) is readily solved
\begin{eqnarray}
&& f_{\bf k}(t)=\int_0^\infty dt' \tilde{\phi}_{\bf k}(t-t')
e^{-\kappa{\bf k}\hat{\Lambda}(t,t-t';t_0){\bf k}},
 \label{Dif6}\\&&
 \hat{\Lambda}(t_1,t_2;t_0)\equiv
\int_{t_2}^{t_1}\hat{W}(\tau,t_0)\hat{W}^T(\tau,t_0) d\tau.
\label{Dif6d}
\end{eqnarray}
Performing inverse Fourier transform of (\ref{Dif6}), one gets
\begin{eqnarray} &&
f(t,{\bf R})=\int_0^\infty\frac{dt'}{4\pi\kappa
\sqrt{det \hat{\Lambda}(t,t-t';t_0)}}\int
d{\bf R}'
\phi\biggl(t-t',\hat{W}^{-1}(t-t',t_0){\bf R}'\biggr)\times\nonumber\\
&&\times\exp\biggl[-\frac{({\bf R}-{\bf R}')\hat{\Lambda}^{-1}(t,t-t';t_0)
({\bf R}-{\bf R}')}{4\kappa}\biggr].
\label{Dif7}
\end{eqnarray}
The auxiliary moment of time $t_0$ may be removed from the solution
after a change of variables $d{\bf R}'\rightarrow d{\bf r}'$,
${\bf r}'= \hat{W}(t-t',t_0)({\bf R}'-{\bf R})$:
\begin{equation}
\theta(t,{\bf r})=\int_0^\infty\frac{dt'}{4\pi\kappa
\sqrt{\det\hat{M}(t,t-t')}}\int d{\bf r}'
\phi\biggl(t-t',{\bf r}'+\hat{W}(t,t-t'){\bf r}\biggr)
\exp\biggl[-\frac{{\bf r}'\hat{M}^{-1}(t,t-t'){\bf r}'}{4\kappa}\biggr],
\label{Dif8}
\end{equation}
where
\begin{equation}
\hat{M}(t_1,t_2)\equiv \hat{\Lambda}(t_1,t_2;t_2)=
\int_{t_2}^{t_1}\hat{W}(\tau,t_2)\hat{W}^T(\tau,t_2) d\tau,
\label{Dif9}
\end{equation}
and we used the following features of $\hat{W}$ as a time-ordered exponent
of a traceless matrix: $\det[\hat{W}]=1$, $\hat{W}^{T}\Lambda^{-1}\hat{W}=
(\hat{W}\hat{\Lambda}(\hat{W}^T)^{-1})^{-1}$.
The expression (\ref{Dif8}) is the formal solution
of (\ref{i6}) and is the direct
generalization of the diffusionless expression (\ref{i8}).

Similarly to what has been done in Sect.I for the diffusionless case,
correlation functions of the passive scalar could now be rewritten in terms of
the
known correlation functions of  the pumping. Performing
integrations with respect to inner temporal (realizing the
source $\delta$-function) and spatial (convolution of two diffusive
Green functions) variables we arrive at the following general expression
\begin{eqnarray}
&&\langle\theta ({\bf r}_1)\theta ({\bf r}_2)\ldots\theta ({\bf
r}_{2n})\rangle=
P_2^n\biggl\langle\sum_{\{ l\}}\prod_{i=1}^{2n}\sqrt{t({\bf r}_{i l_i};\kappa)}
\biggr\rangle_\sigma , \label{Dif11a}\\&&
t({\bf r};\kappa)=\int_0^\infty\frac{dt}{8\pi\kappa
\sqrt{det \hat{M}(0,-t)}}
\int d{\bf r}'\xi_2\biggl(\mid{\bf r}'+\hat{W}(0,-t){\bf r}\mid\biggr)
\exp\biggl[-\frac{{\bf r}'\hat{M}^{-1}(0,-t){\bf r}'}{8\kappa}\biggr]\ ,
\label{Dif11b}
\end{eqnarray}
where $\{ l\}\equiv\{l_1,\ldots,l_{2n}\}$ is a reordering set of numbers
from $1$ to $n$. The term with the maximal
number of integrations (each giving the large logarithmic parameter)
is kept in (\ref{Dif11b}) while the terms with
$P_n$ for $n>2$ are omitted  with logarithmic accuracy like it has been done in
(\ref{U05},\ref{U07}). The expression (\ref{Dif11b}) is thus valid when all the
distances between points $r_{il}$ are much less that $L$ yet maybe however
small.
Let us show that if all the distances are large comparing to $r_{dif}$ then the
further simplification is possible: the term ${\bf r}'$ could be neglected in
the argument of $\xi_2$ so that the integration over $d{\bf r}'$ gives unity
and
we come back to the formalism of Sects. I and III. To see that, we should
compare
typical fluctuations of two terms under the argument
of $\xi_2$. For the largest eigenvalue
of $\hat{M}(0,-t)$ [the smallest eigenvalue of  $\hat{M}^{-1}(0,-t)$],
the eigenvector ${\bbox \rho}_0$
increases exponentially $
\hat{M}(0,-t) {\bbox \rho}_0\sim ({\bbox \rho}_0/\lambda){\exp(2\lambda t)}$,
which gives the following estimation for
a characteristic fluctuation of $ r' $:
$r'\sim \sqrt{{\kappa}/{\lambda}} e^{\lambda t}$.
The second term under the argument of
$\xi_2$ grows with the same exponent $\mid\hat{W}(0,-t){\bf r}\mid\sim  r
e^{\lambda t}$ according to Sect. II.
Therefore, the terms under
the argument of $\xi_2$ differ in prefactors only.
Comparison of the prefactors shows that :\\
1) if $r\gg \sqrt{\kappa/\lambda}$ one can neglect
diffusion coming from (\ref{Dif11b}) back to $t(r;0)$ which has been considered
in Sects. I and III;\\
2) if $r \ll \sqrt{\kappa/\lambda}$ we can drop
${\bf r}$ dependence at all to obtain the
major contribution. If all the distances between points are less that
$\sqrt{\kappa/\lambda}$ the correlation functions are getting independent of
the
distances and could be considered at one point.
The rest of this Appendix is devoted to the second case.

Generally, doing calculations in the way presented above
we see that $\sqrt{\kappa/\lambda}$ (which we named
the diffusion scale $r_{dif}$) is nothing but the ultraviolet cut off which
should be put into diffusionless expressions like (\ref{i10})
at calculating simultaneous correlation functions.
Indeed, the one-point version of (\ref{Dif11a}, \ref{Dif11b}) gives
\begin{eqnarray}
&&\langle\theta^{2n}\rangle=(2n-1)!! P_2^n\langle t^n(0;\kappa)\rangle
\nonumber\\&&=
(2n-1)!! P_2^n\biggl\langle \biggl[\int_0^{\infty}dt
\biggl(\det\biggl[\hat{1}+\hat{M}(0,-t) \frac{8k}{L^2}\biggr]
\biggr)^{-1/2}\biggr]^n
\biggr\rangle_{\sigma},
\label{Dif18}
\end{eqnarray}
where we put $\xi_2(x)$ in the Gaussian form $\exp(-x^2/L^2)$
(recall that with the logarithmic accuracy the final result
(\ref{Dif19}) will be the same for
any function $\xi_2(x)$ if $\xi(0)=1$ and it has the characteristic scale $L$).
The integrand in (\ref{Dif18}) is equal to unity until the moment of time
$t=\ln[L/r_{dif}]/\lambda$, when
both terms under the determinant in (\ref{Dif18})
are of the same order, further in time the integrand decreases exponentially.
That expression for the effective integration time
is exact with the required logarithmic accuracy.
That leads us to the final expression [compare with
(\ref{U077})]
\begin{equation}
\langle\theta^{2n}\rangle=(2n-1)!! \Bigl(P_2 \ln[L/r_{dif}]\Bigr)^n
\langle \lambda^{-n}\rangle_{\sigma}\ .
\label{Dif19}
\end{equation}
We thus expressed the moments of the scalar via the moments of the inverse
stretching rate.
The knowledge of all the moments of $\theta$ (odd ones can be neglected)
allows us to restore the p.d.f. of $\theta$
as an average with respect to the statistics of $\lambda$
\begin{equation}
P(\theta)=\Biggl\langle\sqrt{\frac{\lambda}{2\pi P_2\ln(L/r_{dif})}}
\exp\biggl(-\frac{\theta^2\lambda}{2 P_2 \ln(L/r_{dif})}
\biggr)\Biggr\rangle_{\sigma}.
\label{Dif19a}
\end{equation}

What we learned from Sects. II--IV is that (whatever
be the statistics of the velocity field) the probability distribution of the
stretching rate has Gaussian core and exponential tails --- see
(\ref{Gauss1}--\ref{Gauss3}). Therefore, the same is true for the one-point
statistics of $\theta$ so that for $n<(\bar\lambda/\Delta)\ln Pe$
\begin{equation}
\langle\theta^{2n}\rangle=(2n-1)!! \Bigl(P_2 \ln[L/r_{dif}]\Bigr)^n
{\bar\lambda}^{-n}=(2n-1)!!\langle\theta^{2}\rangle^n \ ,
\label{Dif20}
\end{equation}
That means that $P(\theta)$ also has a Gaussian core. The tails of $P(\theta)$
[determining moments with $n\gg(\bar\lambda/\Delta)\ln Pe$] are exponential as
well as that of $P(\lambda)$. For example, taking $P(\lambda)$ in
the form (\ref{Gauss2}) [with $r_{12}=r_{dif}$], one obtains for
$\ln(L/r_{dif})\gg
\Delta/\bar{\lambda}$
\begin{equation}
P(\theta)\propto \exp\biggl(-\frac{\bar{\lambda}}{\Delta}
\sqrt{\ln^2(L/r_{dif})+\theta^2\Delta/P_2}\biggr).
\label{Dif21}
\end{equation}
We thus conclude that,
{\em all the statements concerning the statistics of the products of the
passive scalar in the
convection interval are true in the diffusion interval as well}.

Besides if one considers the statistics of the differences $\theta({\bf r}_i)-
\theta({\bf r}_j)$ at small enough separations ($r_{ij}\ll r_{dif}$)
then the major (logarithmic) contributions are canceled and only power terms
remain. For example,
\begin{eqnarray}
&&
\langle(\theta ({\bf r}_1)-\theta ({\bf r}_2))^2\rangle =
2 P_2\langle t(0;\kappa) - t(r_{12};\kappa)\rangle=\nonumber\\&&
 P_2
\biggl\langle\int_{0}^{\infty}
\frac{dt }{4\pi\kappa
\sqrt{det \hat{M}(0,-t)}} \,
\int d{\bf r}'\xi_2(r')\exp[-\frac{{\bf r}'\hat{M}^{-1}{\bf
r}'}{8\kappa}]\times
\nonumber\\&&\times
\biggl[1-\exp\biggl(-\frac{{\bf r}\hat{W}^T\hat{M}^{-1}\hat{W}{\bf r}+
{\bf r}\hat{W}^T\hat{M}^{-1}{\bf r}'+{\bf r}'\hat{M}^{-1}\hat{W}{\bf r}}
{8\kappa}\biggr)\biggr]\biggr\rangle_\sigma .
\label{Dif15}
\end{eqnarray}
The integral in the right-hand side of (\ref{Dif15}) stems from the diffusion
region
at small enough times $t\sim r^2_{12}/\kappa\ll 1/\lambda $.
To calculate it at small scales, we can reduce the situation to the pure
diffusion one ($\hat{W}(0,-t)\rightarrow 1, \hat{M}(0,-t)\rightarrow t$)
\begin{equation}
\langle(\theta ({\bf r}_1)-\theta ({\bf r}_2))^2\rangle
\mid_{r_{12}\rightarrow 0}\rightarrow \frac{P_2 r_{12}^2}{4\kappa}\ ,
\label{Dif16}
\end{equation}
which, of course, could also be obtained by the direct integration of
(\ref{flux2}).

\section{Description of the stretching in polar coordinates}
\label{sec:Pc}

This Appendix is devoted to the alternative formalism
based on the representation of (\ref{E1}) in the polar coordinates:
${\bf R}=(R\cos\vartheta,R\sin\vartheta)$:
\begin{equation}
\dot R=\alpha  R\,,\qquad
\dot\vartheta=\beta +c
\label{W03} \,,
\end{equation}
where $\alpha =-a\cos(2\vartheta)-b\sin(2\vartheta)$ and
$\beta = a\sin(2\vartheta)-b\cos(2\vartheta)$.
It is remarkable that the equation for $\vartheta$ is separated,
it can be treated as a constrain enabling us to express
the angle $\vartheta$ via the fields $a,b$. After that is done
the equation (\ref{W03}) for $R$ becomes a scalar equation
with the solution
$ R(t)=r\exp\bigl[ \int_{0}^t dt'\, \alpha (t')\bigr]$.
Already this expression enables us to assert that in the
limit $t\rightarrow\infty$ the statistics of $\ln R$
is Gaussian since the random field $\alpha (t)$ can be shown to have
a finite correlation time. Note that the representations (\ref{W03}) and
(\ref{R2}) could be related by using exponential substitution for $\psi$.

We use this representation to study the non-Gaussian tails
of the p.d.f. $P(t,\lambda)$ for the quantity
$\lambda(t)=\ln[R(t,0)/r]/t$ at small $\lambda$. The p.d.f.
can be written in the following form
\begin{equation}
P(t,\lambda)=t\Biggl\langle \delta\biggl(\lambda t -
\int\limits_0^t dt'\,\alpha \biggr)\Biggr\rangle\equiv
(t/2\pi)\int\limits_{-\infty}^\infty dx
\exp(-ix\lambda t) Y(t,x)
\label{W09} \,,
\end{equation}
where
\begin{equation}
Y(t,x) =Z^{-1}\int {\cal D}\alpha
{\cal D}\beta  {\cal D}c
\exp\biggl[-\int_0^t dt' ({\cal L}-ix\alpha )\biggr]
\label{W10} \,.
\end{equation}
Here $\cal L$ is the Lagrangian density determining the action $S=\int {\cal L}
\,dt$, and $Z$ is the normalization constant so that $Y(t,0)=1$.
In the limit $t\gg\tau$ one gets
$\ln Y\propto t$. First we calculate the contribution
of small $x$ into $P(t,\lambda)$. We can formulate the expansion
of $\ln Y$ in $x$. The first two terms of the expansion are
\begin{equation}
\ln Y(t,x)\simeq ixt\bar\lambda
-tx^2\Delta/2
\label{W11} \,,
\end{equation}
where $\bar\lambda= \langle \alpha \rangle$
is precisely the Lyapunov exponent and $\Delta=\int dt'\langle
\alpha (t')\alpha (t'')\rangle_c$ is the variance.
The angular brackets here designate the average which can
be calculated as the integral over $\alpha , \beta , c$
with the weight $Z\exp(-S)$ and $\langle\dots\rangle_c$
designates the irreducible correlation function. We expect that
(\ref{W11}) determines two first terms of the regular
expansion of $\ln Y$ in $x$ with the convergence radius
of the order of $\bar\lambda/\Delta$. The point
is that the coefficients of this expansion can be expressed
like $\bar\lambda$ and $\Delta$ via the irreducible functions of
$\alpha $, the irreducible function of the $n$-th order
can be estimated as $S^n \tau^{n-1}$ what gives the
convergence radius $(S\tau)^{-1}\sim\bar\lambda/\Delta$.

After substitution of (\ref{W11}) into (\ref{W09}) we
obtain the Gaussian p.d.f.
\begin{equation}
P(t,\lambda)=
\sqrt{t\over 2\pi\Delta}
\exp\biggl(-{(\lambda-\bar\lambda)^2
\over 2 \Delta}t\biggr)
\label{W14} \,.
\end{equation}
The higher in $x$ terms of the expansion of $\ln Y$
(beginning from the third order term) will produce corrections
to the $\ln P(t,\lambda)$ which are small in the
parameter $(\lambda-\bar\lambda)/\bar\lambda$ near
the maximum of $P(t,\lambda)$ but are of the order of
unity for small $\lambda$. Nevertheless the regular character
of the expansion of $\ln Y$ in $x$ ensures that there
are no singular in $\lambda$ terms originating from the
region of integration $x\lesssim\bar\lambda/\Delta$.
Thus this region produces for small $\lambda$
\begin{equation}
P(t,\lambda)\propto
\exp\biggl(-{f(\lambda)\over 2 \Delta}t\biggr)
\label{W15} \,,
\end{equation}
where $f(\lambda)$ is an analytical in $\lambda$ function
which value at small $\lambda$ is
of the order of $\bar\lambda^2$.

The non analyticity of
$P(t,\lambda)$ at small $\lambda$ might be expected from
the region of integration $x\gg \bar\lambda/\Delta$
in (\ref{W09}).
The reason why one may worry about large values of the fields
while studying the probability of anomalously slow stretching
is related to the possibility of suppressing $\lambda$ due to large values
of the field $c$ describing the vorticity. However,
the straightforward analysis presented
in \cite{CFKL} for arbitrary velocity statistics leads
to the conclusion that the region of large $x$ does not
produce a relevant contribution to $P(t,\lambda)$ at small
$\lambda$ and therefore does not change
an exponential behavior of the non-Gaussian tails of $P(t,\lambda)$.

\section{Gaussian velocity field}

The exact expression for the p.d.f. of the stretching time in the case
of delta-correlated velocity field is obtained in Appendix \ref{subsec:DC}.
In Appendix \ref{subsec:FCT}
we use the substitution (\ref{f9}) to prove the statements necessary for
the central limit theorem in the particular case of the Gaussian strain
and to evaluate the correlation time of the stretching rate fluctuations.

\subsection{Statistics of the passive scalar for the delta-correlated
velocity field}
\label{subsec:DC}

In this Appendix we find the {\em exact} expression
for the p.d.f. of $Q=P_2t(r_{12})$
in the case of delta-correlated velocity field.
According to (\ref{Rwr}) one can write
$$ Q=P_2\int_0^T\xi_2\left(R(T-t)\right)dt,$$
where ${\bf R}=\hat{W}(T,t){\bf r}$ and
$T$ is a large value (final answers imply that we take
the limit $T\rightarrow\infty$). We will look for
the Laplacian transform ${\cal P}(s)$ of the p.d.f. for $Q$:
${\cal P}(s)=\langle\exp({-sQ})\rangle$.

Considering delta-correlated initial measure (\ref{i5})
we immediately obtain from (\ref{i73}) that the rotation
does not effect the passive scalar statistics at all. Thus we can
put $c=0$ in the formulas (\ref{f9}-\ref{R2}).
The weight of averaging with respect to
$\psi^{\pm}$ gets the following form
\begin{equation}
{\cal D} \psi^{\pm} \exp\left\{{2\over D}\int_0^T\Bigl[\dot{\psi}^+\psi^--
4(\psi^+\psi^-)^2+2D\psi^+\psi^-\Bigr]dt\right\}\ .
\label{gm}
\end{equation}
We present here some new variant of bosonization procedure that
has been used in the work \cite{Kol1} for $1d$ localization.
First, one makes a gauge transformation
\begin{equation}
\psi^{\pm}(t)=\chi^{\pm}\exp\left(\mp 8\int^T_t \chi^+\chi^-dt'\right),
\label{1s}
\end{equation}
leading to the expression ${R}^2(T-t)=-2r^2\chi^+(t)$.
The initial condition for $\chi^+(t)$ is
$\chi^+(T)=-1/2$ and the transformation Jacobian has the following form
$J[\psi\rightarrow\chi]=\exp(-4\int_0^T\chi^+\chi^-dt)$.
The new integration measure is
\begin{equation}
{\cal D} \chi^{\pm}{\cal D}\rho
\exp\left[-\int_0^T\biggl(-\frac{2}{D}\dot{\chi}^+\chi^--
8\rho\chi^+\chi^-+2D\rho^2\biggr)dt\right],
\label{gm1}
\end{equation}
where on the first step of the bosonization procedure we introduced
the new field $\rho$ by means of the Hubbard-Stratonovich trick.
The second bosonization step is again a gauge transformation
\begin{equation}
\chi^{\pm}(t)=\tilde{\chi}^{\pm}\exp\Biggl(\pm 4D\int^T_t \rho dt'\Biggr) \ ,
\ \ \tilde{\chi}^+(T)=-\frac{1}{2},
\label{2s}
\end{equation}
with the Jacobian $J[\chi\rightarrow\tilde{\chi}]=
\exp(2D\int_0^T\rho dt)$.
A regularization of the transformation (\ref{2s}) as well as (\ref{1s})
are assumed to provide the elimination of nonlinearities in the
suitable discretized expressions (see \cite{ChFyKo}).
As a result we obtain the Gaussian measure
\begin{equation}
{\cal D} \tilde{\chi}^{\pm}{\cal D}\rho
\exp\Biggl[-\int_0^T\biggl(-\frac{2}{D}\dot{\tilde{\chi}}^+\tilde{\chi}^-
+2D\rho^2\biggr)dt\Biggr],
\label{gm2}
\end{equation}
and ${R}^2(T-t)=-2r^2\tilde{\chi}^+(t)\exp(4D\int_t^T\rho dt')$.

An average $\langle F[\tilde{\chi}^+(t)]\rangle$
of an arbitrary functional $F[\tilde{\chi}^+(t)]$
with respect to the measure (\ref{gm2}) is equal to
$F[-1/2]$. This result is easy to get shifting $\tilde{\chi}^+
\rightarrow -1/2+\tilde{\chi}^+$ and noting that the average of
an arbitrary degree of $\tilde{\chi}^+$ is equal to zero. Thus, after the
integration over ${\cal D} \tilde{\chi}^{\pm}$ we arrive at the measure
\begin{equation}
e^{-DT/2}{\cal D} \rho
\exp\left[2D\int_0^T\bigl(\rho-\rho^2\bigr)\,dt\right],
\label{gm3}
\end{equation}
and ${R}^2(T-t)=r^2\exp(4D\int_t^T\rho dt')$. In (\ref{gm3}),
the normalization factor $\exp(-DT/2)$ provides $\langle 1\rangle=1$.
Substituting $\rho=-\dot{\varsigma}, \varsigma(T)=0$, we conclude
that the calculation of $\langle\exp(-sQ)\rangle$ becomes the
quantum mechanical problem with respect to $\varsigma$ variable
\begin{eqnarray}
& &{\cal P}(s)=e^{-DT/2}\int_{\varsigma(T)=0}{\cal D} \varsigma\,
\exp\Biggl\{-\int_0^T\biggl[2D\dot{\varsigma}^2+P_2s
\xi_2\Bigl(e^{2D\varsigma}\sqrt{{r}/{L}}\,\Bigr)+
2D\varsigma(0)\biggr]dt\Biggr\}
\nonumber\\& &=
e^{-DT/2}\Bigl\langle\delta(\varsigma)
|e^{-\hat{H}T}|e^{2D\varsigma}\Bigr\rangle,
\label{qm}
\end{eqnarray}
with the Hamiltonian
\begin{equation}
\hat{H}=-\frac{1}{8D} \partial_{\varsigma}^2+
P_2s\xi_2\Bigl(r\exp(2D\varsigma)\Bigr)
\,.\label{0H}
\end{equation}
The last average in (\ref{qm}) designates a matrix element of
$\exp(-\hat{H}T)$ between states described by the corresponding
wave functions. Let us take for $\xi_2(x)$ the
step function $\vartheta(L-x)$ what will give us the correct
answer in the principal order in $\ln(L/r_{12})$. Then
$\hat{H}=-\frac{1}{8D} \partial_{\varsigma}^2+U(\varsigma)$, where
\begin{equation}
U(\varsigma)=\Biggl\{ \begin{array}{c}
U_s=P_2 s \ , \quad \varsigma<\varsigma_0=\frac{\ln(L/r)}{2D};\\
0 \ , \quad \varsigma>\varsigma_0.\end{array}
\label{U}
\end{equation}
Thus, ${\cal P}(s)$ is equal to $e^{-DT/2}\Psi(\varsigma=0,T)$,
where $\Psi$ is defined from the
following initial value problem: $
\partial_T\Psi=-\hat{H}\Psi$, $\Psi(\varsigma,0)=e^{2D\varsigma}$.
At $\varsigma$ goes to infinity the solution
growing exponentially is proportional to
$e^{2D\varsigma}$ for any $T$.
So, at
$\varsigma\rightarrow +\infty  , \Psi(\varsigma,T)\rightarrow
e^{DT/2+2D\varsigma}$.
In general terms, at $T\rightarrow \infty$ $\Psi(\varsigma)$ approaches
 $e^{DT/2}f(\varsigma)$,
where $f(\varsigma)$ satisfies the equation
$\biggl[\partial_{\varsigma}^2-8DU(\varsigma)-4D^2\biggr]f=0$
and has the asymptotics
$ f(\varsigma\rightarrow +\infty)=e^{2D\varsigma}$ and
$f(\varsigma\rightarrow -\infty) <\infty$.
For the potential (\ref{U}), it gives
$$f(\varsigma)=\Biggl\{ \begin{array}{c}
e^{2D\varsigma}+Ae^{-2D\varsigma} \ , \  \varsigma>\varsigma_0\,;\\
B e^{2D\sqrt{1+2U_s/D}\varsigma} \ , \  \varsigma<\varsigma_0\,.
\end{array}$$
Here, $A$ and $B$ constants have to be defined from matching
$f$ and $\partial_{\varsigma} f$ at the point $\varsigma_0$.
We obtain, finally,
\begin{equation}
{\cal P}(s)=f(\varsigma=0)=B=
\frac{2}{1+\sqrt{1+{2P_2s}/{D}}}\exp\biggl[\ln\frac{L}{r}\Bigl(1-
\sqrt{1+{2P_2s}/{D}}\,\Bigl)\biggl].
\label{Psf}
\end{equation}
This function has the cut along the real axis
from $s^*=-\frac{D}{2P_2}$ to $-\infty$ that gives us
\begin{eqnarray}
& &{\cal P}(Q)=\frac{1}{2\pi i}\int_{-i\infty}^{+i\infty}
e^{sQ} {\cal P}(s)ds=\frac{D}{2P_2}{\cal F}\bigl(Q{D}/{2P_2}\bigr),
\nonumber\\
& &{\cal F}(y)=\frac{2}{\pi} e^{-y+\ln(L/r)}
\int_{-\infty}^{+\infty}\frac{xdx}{1-ix}e^{-yx^2+ix\ln(L/r)}.
\label{Psf1}
\end{eqnarray}
At $y\equiv Q{D}/{2P_2}
\sim \ln(L/r)\gg 1$ the integral in (\ref{Psf1}) calculated
by means of the saddle-point approximation is
\begin{equation}
{\cal F}(y)\approx \frac{2}{\sqrt{\pi y}}\frac{\ln(L/r)}{2y+\ln(L/r)}
\exp\biggl[-\frac{1}{4y}(2y-\ln(L/r))^2\biggr],\label{Psf2}
\end{equation}
that shows the same Gaussian bump as the
formula (\ref{PdfQ}) and non-Gaussian tails discussed in
the Sect. III. Pre-exponential factor in (\ref{Psf2}) is correct for
finite deviations of $y$ from the mean value
(within many dispersion intervals) as long as  $y\gg1$.
For a non-ergodic flow, the averaging over the different
spatial regions may change the form of the probability distribution.
Let us demonstrate that by averaging the p.d.f.
${\cal P}(Q)$ over some smooth
pumping p.d.f. $P(P_2)$. Integrating $P(P_2)$ over $dP_2$
with the function (\ref{Psf1}) one gets in the limit of a large logarithm
$\overline{{\cal P}(Q)}\propto P\bigl[Q\bar\lambda/\ln(L/r_{12})\bigr]$.

\subsection{Finite correlation time of the fluctuations of the stretching rate}
 \label{subsec:FCT}
In this subsection, we show that the fluctuations of the stretching rate have
finite correlation time which is generally does not coincide with $\tau$
and depends on $\bar\lambda$ as well.
Let us suppose that the fields $\varphi^{\pm}(t)$ and $c(t)$
obey the Gaussian statistics with the correlators
\begin{equation}
\langle \varphi^+(t)\varphi^-(t')\rangle=K_1(t-t')\ , \
\langle c(t) c(t')\rangle = K_2(t-t'),
\label{fa1}
\end{equation}
where the functions $K_{1,2}(t)$ correspond to a finite
correlation time $\tau$
\begin{equation}
K_{1,2}(t)\leq const\,e^{-|t|/ \tau}.
\label{fa2}
\end{equation}
In this Appendix we study statistical properties of the quantity
$\Lambda=(\int_t^T (\Lambda_1+2ic) dt'+\Lambda_2(t))/T$,
where $\Lambda_1=4\psi^+\psi^-$ , $\Lambda_2=\ln(-2\psi^+(t))$ and
the equation (\ref{f9}) defines $\Lambda$ as a functional
of the fields $\varphi^{\pm}$ and $c$.
The case of zero correlation time $\tau$ and more general statistics of
the local random fields has been examined exhaustively by H.Furstenberg
\cite{Fur}. The generalization of his results to the case of finite
$\tau$ requires some elaboration because the direct application of a
perturbation theory
leads to infrared divergences. Reformulating
the perturbation theory in a convergent form we prove the stability
of the Furstenberg's results with respect to the finite correlation time.
That is,
we show that with the assumptions (\ref{fa1},\ref{fa2}) the following three
statements are valid:

1. The distribution functional ${\cal P}[\Lambda_1 (t)]$ is
positively defined and differs from zero only for the configurations where
$\Lambda_1(t)\geq 0$ [in other words $\Lambda_1 (t)$ is a nonnegative
random variable].

2. The two-time correlators $D_{i,j}(t_1,t)$ of fluctuations of
$\Lambda_{i,j}(t)$ for $i,j=1,2$
\begin{equation}
D_{i,j}(t_1,t)=\langle \Lambda_i(t_1)\Lambda_j(t)\rangle_c\equiv
\langle\Lambda_i(t_1)\Lambda_j(t)\rangle-\langle\Lambda_i(t_1)\rangle
\langle\Lambda_j(t)\rangle,
\label{fa4}
\end{equation}
go to $0$ at $t_1-t$ goes to $\infty$.

3. The asymptotic relaxation rate of
$D(t_1,t)\equiv \langle \Lambda(t_1)\Lambda(t)\rangle_c$ at the conditions
(\ref{fa1},\ref{fa2}) is limited from below by
$\min\{\bar{\lambda}=\langle \Lambda \rangle, 1/\tau\}$.

To prove the statements 2 and 3 we need to elaborate the
non-linearity of Eq. (\ref{f9}) defining $\Lambda_1$ in terms of
the initial fields $\varphi^{\pm}, c$.
It cannot be treated perturbatively
since it gives spurious time correlations:
the $n$-th order of the perturbative expansion would
have the effective correlation time $\sim n \tau$ going to $\infty$
when $n\to\infty$. The relaxation phenomenon is in this sense
non-perturbative and has essentially dynamical nature.
(The above statement 3 means that the relaxation is governed
by the quantity $<\Lambda>$ which is determined by dynamics).
To avoid a
cumbersome description, we present the complete proof for the strictly
finite correlation time: $K_{1,2}(t)= 0$ if $\quad |t|> \tau.$
The slight modifications required in the more physical case
(\ref{fa2}) are given at the end of this Appendix.

1. This statement is equivalent to the non-negativity of all the
$\Lambda_1(t)$
correlators.
The averages
$\langle \Lambda_1(t_1)...\Lambda_1(t_n)
ic(t'_1)...ic(t'_m)\rangle$
are real due to the time-inversion invariance of the problem.
The Gaussian statistics of the fields
$\varphi^{\pm}(t),c(t)$ allows for the decoupling:
\begin{eqnarray}
& & \langle \Lambda_1(t_1)...\Lambda_1(t_n)\rangle
=\sum_{j_1,...,j_n} \int dt'_1...dt'_n
K_1(t_1-t'_{j_1})...K_n(t_n-t'_{j_n})\times\nonumber\\
& & \biggl[\biggl\langle \frac{\delta \psi^+(t_1)}{\delta \varphi^+(t'_{j_1})}
...\frac{\delta \psi^+(t_n)}{\delta \varphi^+(t'_{j_n})}\biggr\rangle+
...\biggr],
\label{fa8}
\end{eqnarray}
where the set ${j_1,...,j_n}$ runs over all the permutations of numbers
$1,...,n$. The key equality used in this Appendix follows from (\ref{f9}):
\begin{equation}
\frac{\delta \psi^+(t)}{\delta \varphi^+(t')}=
\exp\biggl\{-2\int_t^{t'}\bigl[\Lambda_1(t'')+ic(t'')\bigr]dt''\biggr\}.
\label{fa9}
\end{equation}
Using (\ref{fa9}) we can rewrite (\ref{fa8}) as a linear
combination (with nonnegative coefficients) of averages
\begin{equation}
\biggl\langle
\exp\biggl\{-2\sum_j\int_{t^*_j}^{t^*_{j+1}}
\bigl[\Lambda_1(t')+ic(t')\bigr]dt'
\biggr\}\biggr\rangle,
\label{fa10}
\end{equation}
with some $\{t^*_j\}$. According to the well-known Kubo cumulant formula
every expectation value (\ref{fa10}) can be expressed as an exponential
function of a series in irreducible correlators of the initial exponent.
Their real
values  provide the positivity of the
right-hand side of (\ref{fa8}).

2. First, let us prove that
 \begin{equation}
{\cal G}(t)=\biggl\langle\exp\biggl[-2\int_t^{T}\Lambda_1(t')dt'\biggr]
\biggr\rangle
\rightarrow 0,
\label{fa11}
\end{equation}
when $T-t\rightarrow\infty$. We have already shown that $\Lambda_1(t)$
is nonnegative. Thus, ${\cal G}(t)$ is non growing monotonic
function of $T-t$. The only admissible asymptotic behavior in this
case is ${\cal G}(t)\rightarrow {\cal G}_{\infty}$ where
${\cal G}_{\infty}$ is some finite constant. Let us suppose that
${\cal G}_{\infty}\neq 0$. For any $t_0>t'>t$ the following inequality holds
\begin{equation}
{\cal G}(t_0)\geq\biggl\langle\exp\biggl[-2\int_{t_0}^{T}\Lambda_1(t'')dt''
-2\int_{t}^{t'}\Lambda_1(t'')dt''\biggr]\biggr\rangle\geq {\cal G}(t).
\label{fa12}
\end{equation}
If $T-t_0\rightarrow \infty$ both ${\cal G}(t_0)$ and
${\cal G}(t)$ approach to ${\cal G}(t_{\infty})$.
Thus, the intermediate term in (\ref{fa12}) has the same
limit independent of $t'$. Taking $n$-th derivative of
(\ref{fa12}) with respect to $t'$ we obtain
 \begin{equation}
\biggl\langle\Lambda_1^n(t)\exp\biggl[-2\int_{t_0}^{T}\Lambda_1(t')dt'\biggr]
\biggr\rangle\rightarrow 0,
\label{fa13}
\end{equation}
for any $n>0$ and any $t<t_0$. We show now that the asymptotics
(\ref{fa13}) contradicts to the assumption
${\cal G}_{\infty}\neq0$. Let $t-t_0> 2\tau$.
Then the field $\varphi^-(t)$ does not correlate with
$\varphi^{\pm}(t_1),c(t_1)$ for $t_1\leq t_0$ and the expectation
value in the left-hand side of (\ref{fa13}) for $n=1$ is equal to
 \begin{eqnarray}
&&\biggl\langle\Lambda_1(t)\exp\biggl[-2\int_{t_0}^{T}\Lambda_1(t')dt'\biggr]
\biggr\rangle=\nonumber\\&&=\int_t^T K_1(t-t')
\biggl\langle
\exp\biggl\{-2\int_t^{t'}\bigl[\Lambda_1(t'')+ic(t'')\bigr]dt''-
2\int_{t_0}^{T}\Lambda_1(t'')dt''\biggr\}\biggr\rangle dt'.
\label{fa14}
\end{eqnarray}
Expanding the function $\exp\bigl\{-2\int_t^{t'}
[\Lambda_1(t')+ic(t')]dt'\bigr\}$
in series in $\Lambda_1(t')$ and taking into account the consequence
(\ref{fa13}) of the assumption ${\cal G}_{\infty}\neq 0$ we obtain
for $t_0\rightarrow\infty$
$$\biggl\langle\Lambda_1(t)\exp\biggl[-2\int_{t_0}^{T}\Lambda_1(t')dt'\biggr]
\biggr\rangle\rightarrow \int_t^T K_1(t-t')
\biggl\langle e^{-2i\int_t^{t'}c(t'')dt''}
\exp\biggl[-2\int_{t_0}^T\Lambda_1(t')dt'\biggr]\biggr\rangle dt'\ .$$
Because of the inequalities $t_0-t>2\tau$, $t''<t'<t-\tau$
the field $c(t'')$ does not correlate with
$\Lambda_1(t_1)$ at $t_1>t_0$ and the asymptotic relation
takes the form
\begin{equation}
\biggl\langle\Lambda_1(t)\exp\biggl[-2\int_{t_0}^{T}\Lambda_1(t')dt'\biggr]
\biggr\rangle\rightarrow
{\cal G}_{\infty}\int_t^T K_1(t-t')
 \exp\biggl[-2\int_t^{t'}\int_t^{t'}K_2(t_1-t_2)dt_1dt_2\biggr]\,dt'\ ,
\label{fa16}
\end{equation}
that differs from zero even in the limit $t_0\rightarrow\infty$.
It contradicts to (\ref{fa13}), and the only possibility
is that ${\cal G}_{\infty}=0$.
Turning back to the correlator $D(t_1,t)$ we note that
(\ref{f9}) is equivalent to the integral equation
\begin{equation}
\psi^+(t)=\psi^+(t_0)\exp\biggl\{-\int_t^{t_0}
\bigl[\Lambda_1(t')+ic(t')\bigr]dt'
\biggr\}+
\int_t^{t_0} \varphi^+(t')\exp\biggl[-\int_t^{t'}(\Lambda_1+ic)dt''\biggr]dt'.
\label{fa17}
\end{equation}
Thus $\psi^+(t)$ depends on the fields $\varphi^{\pm}(t'),c(t')$
at $t'>t_0$ via $\psi^+(t_0)$ only. For every two functionals
$A[\varphi]$ and $B[\varphi]$ of random fields $\varphi(t)$
obeying Gaussian statistics with the correlator
$\langle \varphi(t)\varphi(t')\rangle=K(t-t')$
(indices are assumed) the following equality holds
\begin{eqnarray}&&
\langle A[\varphi]B[\varphi]\rangle_c=\nonumber\\&&
\sum_{n=1}^{\infty}
\frac{1}{n!}\int dt_1...dt_n dt'_1...dt'_n
\biggl\langle \frac{\delta^{(n)}A}{\delta \varphi(t_1)...\delta \varphi(t_n)}
\biggr\rangle K(t_1-t'_1)...K(t_n-t'_n)\biggl\langle
\frac{\delta^{(n)}B}{\delta \varphi(t'_1)...\delta \varphi(t'_n)}
\biggr\rangle .
\label{fa18}
\end{eqnarray}
Let us substitute into (\ref{fa18}) $\Lambda_1(t)$ and $\Lambda_1(t')$
instead of $A[\varphi]$ and $B[\varphi]$.
If $t'-t_0>2\tau$ and $t_0-t>2\tau$ the resulting expressions
incorporate only the functional derivatives
$\delta^{(n)} \Lambda_i(t)/(\delta\varphi(t_1)...\delta\varphi(t_n))$
with $t_k>t_0, k=1,...,n$. Then from (\ref{fa17}) we obtain
\begin{equation}
\frac{\delta^{(n)} \Lambda_1(t)}
{\delta\varphi(t_1)...\delta\varphi(t_n)}=
\frac{\delta^{(n)} \Lambda_1(t)}{(\delta \psi^+(t_0))^n}
\frac{\delta^{(n)} \psi^+(t_0)}
{\delta\varphi(t_1)...\delta\varphi(t_n)},
\label{fa19}
\end{equation}
Here $\varphi(t_j)$ designates $\varphi^{\pm}(t_j)$
or $c(t_j)$.
All the derivatives of $\Lambda_1(t)$ with respect to $\psi^+(t_0)$
can be expressed explicitly in terms of $\Lambda_1(t')$ and the fields
$\varphi^{\pm}$ and $c$. For example,
\begin{equation}
\frac{\delta \Lambda_1(t)}{\delta \psi^+(t_0)}=
4\varphi^-(t)\exp\biggl[-2\int_t^{t_0} (\Lambda_1+ic)dt'\biggr].
\label{fa20}
\end{equation}
Substituting (\ref{fa19}) together with (\ref{fa20}) into
(\ref{fa18}) we see that the irreducible correlators $D_{1,1}(t,t')$
for $t-t'\gg \tau$ can be estimated as
\begin{equation}
D_{1,1}(t,t_1)\leq {\cal G}(t) F(t_0,t_1),
\label{fa21}
\end{equation}
where $F(t_0,t_1)$ are some finite functions of $t_1$ and of fixed
intermediate time moment $t_0$. The behavior (\ref{fa11})
provides for the relaxation of $D_{1,1}(t,t_1)$
\begin{equation}
D_{1,1}(t,t_1)\rightarrow 0, \ \text{at} \  t_1-t\rightarrow \infty.
\label{fa22}
\end{equation}
The relation (\ref{fa17}) together with the proved applicability
of the central limit theorem to the quantity $\int\Lambda_1(t)dt$
gives for the correlator $D_{2,j}(t,t_1)$, $j=1,2$
the asymptotic inequality of the form (\ref{fa21}) with some
another function $F(t_0,t_1)$ in the right hand site.

3. The fact that $\Lambda$ is
self-averaging
gives us the asymptotic behavior of ${\cal G}(t)$ at
$T-t\rightarrow\infty$
\begin{equation}
{\cal G}(t)\rightarrow const \times\exp[-2(T-t)\bar{\lambda}] \ .
\label{fa23}
\end{equation}
{}From (\ref{fa21}) we conclude that in the case of strictly finite
correlation time  the relaxation rate of fluctuations
of $\Lambda(t)$ is bounded from below by $2\bar{\lambda}$.
Let us turn to the case of exponentially decaying correlators
$K_{1,2}(t)$ (\ref{fa2}). Replacing the inequalities like $t-t_0>2\tau$
by  $t-t_0\gg\tau$ and neglecting then the exponentially small
terms $\sim\exp(-(t-t_0)/\tau)$ we make the proofs of the statements
1 and 2 valid in this case as well. In the proof of the statement 3
the integration over intermediate times in (\ref{fa18}) goes
over whole time interval and there will be the contribution
into $D(t,t')$ proportional to $\exp(-(t'-t)/\tau)$.
Such contributions exist already in the zero order of perturbation
theory and arise from the dependence of $\Lambda(t)$ on the fields
$\varphi^{\pm},c$ in the vicinity of the same time moment $t$ when
dynamics is yet linear.
If $(\tau)^{-1}>2\bar{\lambda}$ the relaxation rate of
$D(t,t')$ is again $2\bar{\lambda}$, otherwise
$D(t,t')$ decays with the exponent $(\tau)^{-1}$.

\begin{figure}[htbp]
\begin{center}
\setlength{\epsfxsize}{130mm}
\leavevmode
\epsffile{fig1.ps}
\caption{Lyapunov exponent $\bar\lambda$ as a function of
$\tau$ at $D_s=\tau$. The three plots correspond to different
values of $D_a/D_s$: (a) $D_a=0$; (b) $D_a=D_s$;
(c) $D_a=25 D_s$, the dashed lines marking the asymptotics
that have been obtained analytically.}
\end{center}
\end{figure}

\begin{figure}[htbp]
\begin{center}
\setlength{\epsfxsize}{130mm}
\leavevmode
\epsffile{fig2.ps}
\caption{Lyapunov exponent $\bar\lambda$ as a function of
$\varepsilon =(D_a/D_s)^{1/2}$ at $D_s=\tau =2.$}
\end{center}
\end{figure}
\end{document}